\documentclass[times,doublespace]{qjrms4}

\usepackage[colorlinks,bookmarksopen,bookmarksnumbered,citecolor=red,urlcolor=red]{hyperref}

\newcommand\BibTeX{{\rmfamily B\kern-.05em \textsc{i\kern-.025em b}\kern-.08em
		T\kern-.1667em\lower.7ex\hbox{E}\kern-.125emX}}

\usepackage{moreverb}
\usepackage{graphicx}
\usepackage{xcolor,colortbl}
\usepackage[switch,displaymath,mathlines,modulo]{lineno}
\usepackage{linenofix}
\usepackage{amsmath}
\usepackage[titletoc,toc,title]{appendix}

\def\volumenumber{00}

\newcommand{\be}{\begin{equation}}
\newcommand{\ee}{\end{equation}}
\newcommand{\bmath}{\begin{displaymath}}
\newcommand{\emath}{\end{displaymath}}

\newcommand{\correction}{\color{black}} 

\begin{document}
	
	\runningheads{Metref et al.}{Model evidence using data assimilation with localization }
	
	\title{Estimating model evidence using ensemble-based data assimilation with localization - The model selection problem}
	
	\author{S.~Metref\corrauth\affil{a}, A.~Hannart\affil{a}, J.~Ruiz\affil{a,b}, M.~Bocquet\affil{c}, A.~Carrassi\affil{d} and M.~Ghil\affil{e,f}}
	
	\address{
		\affilnum{a}IFAECI, CNRS-CONICET-UBA, Buenos Aires, Argentina.\\
		\affilnum{b} CIMA-CONICET-University of Buenos Aires.\\
		\affilnum{c}CEREA, joint laboratory \'Ecole des Ponts ParisTech and EDF R\&D, Universit\'e Paris-Est, Champs-sur-Marne, France.\\
		\affilnum{d}Nansen Environmental and Remote Sensing Center, Bergen, Norway.\\
		\affilnum{e}  Geosciences Department and Laboratoire de M\'et\'eorologie Dynamique (CNRS and IPSL), \\ \'Ecole Normale Sup\'erieure and PSL Research University, Paris, France.\\
		\affilnum{f} Dept. of Atmospheric and Oceanic Sciences, University of California, Los Angeles, USA.
		}
	
	\corraddr{ Sammy Metref, 
		IFAECI, CNRS-CONICET-UBA,
        Pab. II, piso 2, Ciudad Universitaria 1428 Buenos Aires, Argentina,
        Tel.: +5411-4787-2693,
        Fax: +5411-4788-3572,
        e-mail: sammy.metref@cima.fcen.uba.ar}
	
	\begin{abstract} 
		In recent years, there has been a growing interest in applying data assimilation (DA) methods, originally designed for state estimation, to the model selection problem. {\correction In this setting,}  \cite{ carrassi17} introduced the contextual formulation of model evidence (CME) and showed that CME can be efficiently computed using a hierarchy of ensemble-based DA procedures. 
Although \cite{ carrassi17} analyzed the DA methods most commonly used for operational atmospheric and oceanic prediction worldwide, they did not study these methods in conjunction with localization to a specific domain. Yet any application of ensemble DA methods to realistic geophysical models requires the implementation of {\correction some form of} localization. 
The present study extends the theory for estimating CME to ensemble DA methods with domain localization. 
The domain-localized CME (DL-CME) developed herein is tested for model selection with two models: (i) the Lorenz 40-variable mid-latitude atmospheric dynamics model (L95); and (ii) the simplified global atmospheric SPEEDY model. 
The CME is compared to the root-mean-square-error (RMSE) as a metric for model selection. The experiments show that CME improves systematically over the RMSE, and that {\correction this skill improvement} is further enhanced by applying localization in the estimate of the CME, using the DL-CME. The potential use and range of applications of the CME and DL-CME as a model selection metric are also discussed. 
	\end{abstract}
	
	\maketitle
	  
	  
	\section{Introduction and motivations}   
	 
	Model selection {\correction in the broader literature is the task of selecting a statistical model from a set of candidate models, given data \cite[e.g.,][]{Ando10}. More precisely \citep{baum70}, it} is the problem of defining and using metrics to compare different models or {\correction different model versions, i.e.,} different versions of the same model that may differ by the model's parameters, its {\correction subgrid-scale parameterisations} or its boundary conditions.  
	
	 {\correction Either model selection or, more specifically, systematic model version selection}, though, have received little attention from the geoscientific community, as compared to data assimilation (DA), which is by now a well-established field in the geosciences (\citealp{daley91}; \citealp{ghil91}; \citealp{bennett92}; \citealp{kalnay03}; \citealp{asch16}; \citealp{carrassi18}). {\correction DA, though,} has focused so far mainly on estimating high-dimensional states of the atmosphere (\citealp{whitaker08}; \citealp{buehner10}), of the ocean (\citealp{lermusiaux06}; \citealp{sakov12}) and of the climate system as a whole (\citealp{saha10}; \citealp{saha14}). Even nowadays, common practice {\correction in evaluating models and successive versions thereof} at operational centers is to compare visually model outputs to observations or, more quantitatively, to compute the model-to-data root-mean-square error (RMSE). 
	 
 Some attempts to analytically address the {\correction model or model version selection} problem have been successful (e.g., \citealp{winiarek11}) but only under specific assumptions that may not be realistic for large-dimensional geoscientific applications.
 Numerical methodologies that may be computationally affordable for high-dimensional problems have also been proposed in recent years (\citealp{sarkka13,elsheikh14a,elsheikh14b,otsuka15,reich15,carson17, carrassi17}). Several of these approaches advocate the use of the marginal likelihood of the data --- also called model evidence --- for model selection.  
 Along these lines, the contextual formulation of model evidence (CME), first proposed by \cite{hannart16b} and fleshed out by \cite{ carrassi17}, exploits DA techniques to approach the model selection problem.  
	 
	 \cite{ carrassi17} provide the analytic formulae to compute the CME using several state-of-the-art Gaussian ensemble DA methods and demonstrate the benefits of CME as a model selecting tool. 
	Their study thus started to answer the first theoretical and practical research questions that had to be addressed to make the CME suitable for {\correction model version selection} operationally in weather forecasting centers.  
	
  Nevertheless, a number of issues remain to be tackled in order to proceed to the operational implementation of the CME.
  Among these, one key issue is the use of spatial localization --- {\correction simply} called henceforth localization --- in ensemble-based DA.
  Indeed, when applied to large-dimensional models, the ensemble based DA methods of \cite{carrassi17} to compute CME would fail unless localization is properly incorporated.  
  Even though localization is a widely established practice in ensemble DA, and has also generated a substantial literature (e.g., \citealp{houtekamer01,hamill01,ott04} and \citealp{sakov11}), it is still {\correction a topic of very active} research. 
	
	Localization originally arose as a countermeasure to the sampling issue due to our attempt to describe uncertainty in high dimensions using a very limited ensemble. 
	Help in overcoming this problem comes from the fact that many physical systems, and notably the atmosphere and ocean, tend to have small long-distance correlations, i.e., the actual error covariances are sparse (e.g., \citealp{ghil79, balg83}).  
	For such systems, localization can -- without degrading too severely a physical system's representation -- set long-distance correlations to zero and only model the terms in the covariance matrix that are close to the diagonal.  

  Many localization techniques exist to achieve sufficiently accurate approximations of the covariance matrix along these lines. 
  These techniques, though, do lead to a change in the structure of the matrices manipulated during the DA process. 
  Hence, in the CME context, the challenge is to adapt its original formulation to the need for localization in the underlying DA process. 
  This study aims, therefore, to extend CME theory, as presented in \cite{carrassi17}, to include localization and to apply it to a quasi-realistic atmospheric model of intermediate complexity.

	The structure of the paper is as follow. 
	Section \ref{ModEvidAndDA} recalls the benefits of approaching model evidence using the DA framework.
	A more extensive discussion of these benefits can be found in \cite{hannart16b}.
	Section \ref{CMEAndLoc} first introduces the notion of localization in general, and the domain localization technique in particular. 
	The application of domain localization in the CME computation is hereafter referred to as the domain-localized CME (DL-CME), and it is also presented in section \ref{CMEAndLoc}. 
	Section \ref{NumExpL95} implements the two CME formulations, with and without localization, and compares them to the RMSE as {\correction indicators of model version selection} in various numerical experiments based on the 40-dimensional L95 model of \cite{lorenz95}.
	Section \ref{NumExpSPEEDY} applies these indicators to a parameter selection problem using the simplified global atmospheric model called SPEEDY (\citealp{molteni03,kucharski06}). 
	Section \ref{RemAndConclu} finally provides a summary, conclusions and future directions. 
	 
	 
	\section{Model evidence and data assimilation}   
	\label{ModEvidAndDA} 
	Let us assume that a model $\mathcal{M}$ describes an unknown process at time $k$, as a $M$-dimensional state $\textbf{x}_{k}$, that we only perceive via a set of partial, $d$-dimensional observations $\textbf{y}_{k}$.

	\paragraph{The model $\mathcal{M}$.} The model $\mathcal{M}$ simulates the unknown process
		\begin{equation}
		\textbf{x}_{k}=\mathcal{M}_{k:k-1}(\textbf{x}_{k-1})+\boldsymbol{\eta}_k,
		\end{equation}
		where $\mathcal{M}_{k:k-1}: \mathbb{R}^M\rightarrow\mathbb{R}^M$ propagates the state $\textbf{x}\in\mathbb{R}^M$ from time $t_{k-1}$ to time $t_k$.  
		The dimension $M$ of the model state can be decomposed as 
		$M=\mathrm{Card}(\Lambda)\times\mathrm{Card}(\Gamma)$, 
		where $\Lambda$ is the space of the physical variables, {\correction such as temperature and velocity components, while $\Gamma$ is the set of finite modes of the model that represent the physical variables}. 

		 In general, $\boldsymbol{\eta}_k \in \mathbb{R}^M$ is the model error --- represented most often
		by an additive stochastic white noise --- but here we shall assume that the model is perfect, i.e., $\boldsymbol{\eta}_k\equiv 0$. 
		 The implications of this assumption on the computation of the CME are substantial, cf.~\cite{carrassi17}, but we have reason to believe that it does not affect very much our conclusions about the impact of localization on the CME and its comparison with the RMSE. . 
		 This perfect model assumption will be removed in subsequent studies.  		
		 
		\paragraph{The set of observations $\textbf{y}$.}  
		The observation vector $\textbf{y}_k \in \mathbb{R}^d$ is related to the state vector $\textbf{x}_k$ according to the observation model
		\begin{equation}
			\textbf{y}_k=\mathcal{H}_k(\textbf{x}_k)+\boldsymbol{\epsilon}_k.
			\label{Eq_SetofObs}
		\end{equation}
		Here $\mathcal{H}_k: \mathbb{R}^M\rightarrow\mathbb{R}^d$ is the observational operator at time $k$ and $\boldsymbol{\epsilon}_k$ is the observation error, represented here as an additive stochastic white noise with zero mean and covariance $\mathbf{R}$, which is assumed to be constant in time.  
		 
		Having introduced the observational model, ${\mathcal H}$, in Eq.~\ref{Eq_SetofObs}, we can now define its liberation around ${\bf x}_k$, to be used in the DA process (cf. section~\ref{CMEEnKFForm}), as the $d\times M$ matrix ${\bf H}_k$, and the associated transpose ${\bf H}_k^{\rm T}$.
	 
	\subsection{CME for {\correction model version selection}}  
	 
	Using a {\correction given} model of type $\mathcal{M}$ and the set of observations $\textbf{y}_{k:1}=\{ \textbf{y}_{k},\textbf{y}_{k-1},...,\textbf{y}_{2},\textbf{y}_{1} \}$, the goal of {\correction model version selection} is to define a metric based on them, and apply it to identify the best model.   
	Several metrics for {\correction model version selection} are defined in the literature (\citealp{akaike74,schwarz78,burnham02}). 
	In the present article, we focus on model evidence as the metric of choice (\citealp{sarkka13,elsheikh14a,elsheikh14b,otsuka15,reich15,carson17, carrassi17}). 
	
	\subsubsection{Model evidence}
	
	Let us define the ideal infinite set of observations as $\textbf{y}_{k:}=\{ \textbf{y}_{k},\textbf{y}_{k-1},...,\textbf{y}_{1},\textbf{y}_{0},...,\textbf{y}_{-\infty} \}$.
	The marginal likelihood of $\textbf{y}_{k:}$, given $\mathcal{M}$, is calculated as
	\begin{equation}\label{equ:marge}
	p(\textbf{y}_{k:}|\mathcal{M})=\int\!\! \mathrm{d}\textbf{x}\,\, p(\textbf{y}_{k:}| \textbf{x},\mathcal{M})p(\textbf{x}).
	\end{equation}
	This likelihood is also referred to as {model evidence}.

	Note that, in the context of model evaluation, one should rather be interested in estimating the probability of the model conditioned on the observations $p(\mathcal{M}|\textbf{y}_{k:})$.
	Yet, model evidence can be used to construct a metric {\correction either for model version selection or for the comparison between two distinct} models $\mathcal{M}_0$ and $\mathcal{M}_1$.
	
	To see this, let us write the ratio of the two posterior distributions, each one relative to models $\mathcal{M}_0$ and $\mathcal{M}_1$ respectively, conditioned on the same set of observations 
	\begin{equation}
	\frac{p(\mathcal{M}_1|\textbf{y}_{k:})}{p(\mathcal{M}_0|\textbf{y}_{k:})} = \frac{p(\textbf{y}_{k:}|\mathcal{M}_1)}{p(\textbf{y}_{k:}|\mathcal{M}_0)} \frac{p(\mathcal{M}_1)}{p(\mathcal{M}_0)}
	\label{BayesianRatios}
	\end{equation}
	 
	Equation \eqref{BayesianRatios} computes the ratio of the posterior model probability density functions (pdfs) and we
	see that it is proportional to the ratio of the model evidences -- the Bayesian factor -- times
	the ratio of the prior model pdfs. 
	The latter are difficult to evaluate in practice, but, it is often legitimate to assume that the prior pdfs of the two models are the same, provided that we do not possess enough
	information to favor one model over the other.
	In this case, we see how the Bayesian factor provides an estimate of the model pdf ratio.
	In any case, the Bayesian factor is the scalar that
	updates the ratio of prior model pdfs to the ratio of posterior model pdfs. 
	
	In this sense, the ratio of the two model evidences $p(\textbf{y}_{k:}|\mathcal{M}_1)/p(\textbf{y}_{k:}|\mathcal{M}_0)$ can be considered as a selection indicator that favors one model or the other, depending on $p(\textbf{y}_{k:}|\mathcal{M}_1)/p(\textbf{y}_{k:}|\mathcal{M}_0)\lessgtr1$.
	More conveniently, one can look at the difference between their respective logarithms, and write the model selection indicator as
	\begin{equation}
	\Delta_{\textbf{y}_{k:}}(\mathcal{M}_0,\mathcal{M}_1)=\ln\{ p(\textbf{y}_{k:}|\mathcal{M}_1) \}-\ln\{p(\textbf{y}_{k:}|\mathcal{M}_0)\}.
	\end{equation}
	We use the indicator $\Delta_{\textbf{y}_{k:}}(\mathcal{M}_0,\mathcal{M}_1)$ in the following experiments, and the model selection criterion 
	becomes $\Delta_{\textbf{y}_{k:}}(\mathcal{M}_0,\mathcal{M}_1) \lessgtr 0$.
	 
   Previously, model evidence calculations were mainly based on Monte Carlo methods, as in \cite{elsheikh14a}, \cite{elsheikh14b} and \cite{carson17}.  
   In the present article, we focus on the use of DA for this purpose, following and expanding the work in \cite{ carrassi17}.

		\subsubsection{Contextual model evidence {\correction (CME)}} 
		 
		If the model is autonomous, one can condition model evidence {\correction on} the initial observation $\textbf{y}_0$ of the set, and write
		\begin{equation}
			p(\textbf{y}_{k:}| \mathcal{M})=p(\textbf{y}_{k:1}| \textbf{y}_{0:})p(\textbf{y}_{0:})
		\end{equation}
		 where the explicit dependence on $\mathcal{M}$ has been dropped on the right-hand side for the sake of clarity, and  where $p(\textbf{y}_{k:1}| \textbf{y}_{0:})$ is the likelihood of the observations from time $t_1$ to $t_k$ conditioned on all observations up to and including time $t_0$.
		 
		{\correction 
		A significant difference {\correction between the CME method employed here and previous methods \citep{sarkka13, reich15, delmoral04}}
		 is the use of the current-time density as a prior. 
		This choice allows one to narrow the probability density so as to focus on plausible states given the current conditions of {\correction a system.}
		The latter are embedded in our, inevitably approximate, knowledge of the prior.}
		Indeed, an alternative formulation of the model evidence, which also can give a computational gain, is to no longer try to estimate the climatological model evidence $p(\textbf{y}_{k:}|\mathcal{M})$ -- which is difficult to estimate accurately and which does not provide information on the present context of the system -- {\correction but} the so called contextual model evidence (CME), $p(\textbf{y}_{k:1}| \textbf{y}_{0:})$ \citep{carrassi17}.  
		This new contextual view of model evidence implicitly assumes that all information from past observations is conveyed by conditioning on $\textbf{y}_{0:}$.  

		Marginalizing over $\textbf{x}_0$ yields
		\begin{equation}
			p(\textbf{y}_{k:1}| \textbf{y}_{0:})=\int\!\! \mathrm{d}\textbf{x}_0\,\, p(\textbf{y}_{k:1}| \textbf{x}_0) p(\textbf{x}_0| \textbf{y}_{0:})
			\label{eq_CMEorig}
		\end{equation} 
	Therefore, the two terms needed to compute the CME are: 
	\begin{itemize}
		\item the conditional pdf, $p(\textbf{x}_0| \textbf{y}_{0:})$, and
		\item the likelihood of the observations, $p(\textbf{y}_{k:1}| \textbf{x}_0)$.
	\end{itemize} 
	The conditional pdf $p(\textbf{x}_0| \textbf{y}_{0:})$ is the posterior pdf, or {analysis}, produced by the state estimation DA process. 
	The observational likelihood $p(\textbf{y}_{k:1}| \textbf{x}_0)$ is often a given quantity (or at least considered as such) of the state estimation DA problem. 
	Hence, as shown in \cite{carrassi17}, the CME can be obtained as a by-product of a DA algorithm  designed to assimilate the data $\textbf{y}$ into the model $\mathcal{M}$.

	\subsection{The CME's EnKF-formulation}   
	\label{CMEEnKFForm}
	
	We present here the CME formulation based on the ensemble Kalman filter (EnKF: \citealp{evensen09}). 
	The choice of the EnKF among other, possibly even more accurate, methods to compute CME, is motivated by its great cost/benefit ratio. 
	\cite{carrassi17} showed that
	{\correction the CME evaluated by the EnKF achieves high accuracy and discrimination skills -- that is 
	only slightly lower than when evaluated by other more sophisticated DA methods -- but does so at a much lower computational cost.}
	{\correction
		In fact, \cite{carrassi17} also showed that -- although the DA schemes with the {\correction coarsest} approximations (e.g., the EnKF) provide the worst estimates of the CME in a strongly nonlinear system, {\correction such as the \cite{lorenz63} model -- the performance of these schemes} in estimating the CME is comparable, regardless of their respective approximations, in a weakly nonlinear system like the L95 model. 
		Hence, in {\correction atmospheric, oceanic or coupled} systems at synoptic scales, which behave in a quasi-linear and quasi-Gaussian manner {\correction most of the time,} the differences produced by the choice of assimilation scheme are expected to be negligible compared to the difference between two model {\correction versions.}        
		}
	
	The EnKF-formulation of the CME comes as a by-product of the EnKF DA process that gives an approximation of model evidence $p(\textbf{y}_{K:1}| \textbf{y}_0)$. 
	{\correction Over the evidencing window $\{t_1, ..., t_K\}$, of size $K\in\mathbb{Z}^+$}, 
	the CME is calculated as
	
	\begin{align}
	\begin{split}  
	p(\textbf{y}_{K:1}|\textbf{y}_{0:})&\approx\prod_{k=0}^K (2\pi)^{-\frac{d}{2}}|\boldsymbol{\Sigma}_k|^{-\frac{1}{2}} \\ 
	& \!\!\!\!\!\!\!\!\! \times \exp\left\{-\frac{1}{2}[\textbf{y}_k-\mathcal{H}_k(\textbf{x}^\mathrm{f}_k)]^\mathrm{T}\boldsymbol{\Sigma}_k^{-1}[\textbf{y}_k-\mathcal{H}_k(\textbf{x}^\mathrm{f}_k)]\right\},
	\label{EnKFLikelihoodFormulation}
	\end{split} 
	\end{align}  
	{\correction where $\textbf{x}^\mathrm{f}_k$ is the prior state estimate at time $k$, and where} $\boldsymbol{\Sigma}_k=\textbf{H}_k\textbf{P}^\mathrm{f}_k\textbf{H}_k^\mathrm{T}+\textbf{R}$, while $\mathbf{P}^\mathrm{f}_k$ is the prior error covariance matrix. 
	
	{\correction 
		Note that, whilst the RMSE is the Euclidean distance between the first guess and the observations, the CME can be seen as a distance between the same two {\correction fields,} ponderated by the matrix $\boldsymbol{\Sigma}_k^{-1}$. 
		Hence, the {\correction uncertainty $\textbf{P}^\mathrm{f}_k$ on the prior and the uncertainty $\textbf{R}$ on the data} both penalize the CME. 
		In other words, {\correction overfitting the model to the data, due to a faulty assimilation, will not erroneously} decrease the CME. {\correction
		The CME will thus} be able to spot and penalize a model-to-data overfit, whereas the RMSE that does not take into account the prior and the data uncertainties will not.   
		}
  
	{\correction There are two} different proofs of the EnKF-based CME in Eq.~(\ref{EnKFLikelihoodFormulation}),  
	{\correction one for the stochastic EnKF in \cite{hannart16b} and another for the deterministic EnKF in \cite{ carrassi17}: both are referred to hereafter as the global CME (G-CME).  }

	An efficient alternative to compute the G-CME is presented in \nameref{Appendix1} and it is used throughout the present article.   
	This implementation {\correction assumes} that $\textbf{R}$ is diagonal, i.e., that the observations are mutually uncorrelated, an assumption {\correction that is} often made in the geosciences {\correction and elsewhere.
	A significant improvement achieved by} this alternative G-CME implementation is 
	{\correction that the problem is no longer solved in the physical space {\correction of dimension $M$ but in the reduced ensemble space of dimension $N$, with $N\ll M$ in most cases.} 
	This implementation {\correction thus reduced significantly the computational cost of} the experiments in the present paper.  
	 }

	
	\section{CME and domain localization} 
	\label{CMEAndLoc}
		  This section presents a new formulation of the CME 
		  {\correction with domain localization. }
		  We describe first, in section~\ref{DomLoc}, how domain localization is implemented within the DA process, and then the {domain-localized} CME (DL-CME) in section~\ref{LocaLikeli}. 

			\subsection{Domain localization}
			\label{DomLoc} 
			\subsubsection{The basic idea}  
			
			{\correction 
				Domain localization dates back to earlier forms of DA, like the successive correction method (\citealp{cressman59,ghil79,ghil91}). 
				A variant known as local analysis was then introduced 
				{\correction by \cite{houtekamer01} and by \cite{ott02,ott04}.}
				Here, we describe a different approach to local analysis due to \cite{hunt07} and applicable in {\correction the Ensemble Transform Kalman Filter (ETKF) of \cite{bishop01} framework. The Local ETKF (LETKF) method of \cite{hunt07} consists in} updating separately, at each spatial gridpoint $s \in \Gamma$, all the physical variables $\{ \textbf{x}_{s, l}\}$, with $l \in \Lambda$, by assimilating only neighboring observations; {\correction see section \ref{LETKFmethodo}.} 
				}
			 
		The first step of local analysis consists of selecting the observations within a disk centered on the spatial gridpoint $s\in\Gamma$ we wish to update; the radius of this disk is a predefined parameter $\rho_{\mathrm{cut}}$, called the cut-off radius.
		This type of selection process is often called the box-car scheme, as opposed to schemes that gradually taper off the weights given to observations, from $1$ to $0$. 
		Note that in general the cut-off radius depends on the physical variable to be analyzed. 
  
		We {\correction denote} the resulting reduced observation vector by $\textbf{y}_{|s}$, and its size by $\widetilde{d}$. 
		It is then also necessary to restrict the observation error covariance matrix $\textbf{R}$ to a {\correction local matrix $\textbf{R}_{|s}$ that has smaller dimension. 
		However, the analysis resulting} after the first step alone may {\correction -- when the observations are sparse, for instance (\citealp{szunyogh08}) --}  
		present abrupt discontinuities at the boundary of the observation disk and have, therefore, a deleterious side-effect, whose elimination or mitigation requires a second analysis step. 
			
		The second step in local analysis applies a localization function to the inverse matrix $\textbf{R}_{|s}^{-1}$, called precision matrix, and it produces a smoother global update. 
		The localization function is in fact the Schur product \citep{hunt07} of $\textbf{R}_{|s}^{-1}$ by the diagonal localization matrix $\mathbf{L}$ giving 
		 \begin{equation}
		 	\widetilde{\textbf{R}}_{|s}^{-1}=\mathbf{L}\circ\textbf{R}_{|s}^{-1}=( \textbf{R}_{|s}^{-1})_{i,j} \cdot (\mathbf{L})_{i,j}
		 \end{equation} 
		 where $(\mathbf{L})_{i,i}$ is equal to $1$ if $i=s$ and it decreases gradually to $0$ outside of the disk of radius $\rho_\mathrm{loc}$, called the localization radius. 
		 Hence, the further an observation lies from $s$, the more its observation error variance is increased and its impact on the local analysis is decreased. 
			
			\subsubsection{The LETKF}
			\label{LETKFmethodo}
			
			The ensemble transform Kalman filter (ETKF: \citealp{bishop01}) is a deterministic EnKF performing the analysis in the ensemble space. 
			The ETKF computes the updated ensemble $\textbf{E}^\mathrm{a}=[\textbf{x}^\mathrm{a}_1,...,\textbf{x}^\mathrm{a}_N]\in {\mathrm R}^{M\times N}$, with $\textbf{x}^\mathrm{a}_i$ the $N$ ensemble members, such that
			\begin{equation}
			\textbf{E}^\mathrm{a}=\bar{\textbf{x}}^\mathrm{f} \textbf{1}^\mathrm{T}+\textbf{X}^\mathrm{f}\left(\textbf{w}^\mathrm{a}\textbf{1}^\mathrm{T}+\,\sqrt{N-1}\widetilde{\textbf{P}^\mathrm{a}}^{1/2}\right).
			\end{equation}
			Here 
			$\widetilde{\textbf{P}^\mathrm{a}}=\{ (\textbf{HX}^\mathrm{f})^\mathrm{T}\textbf{R}^{-1}(\textbf{HX}^\mathrm{f})+(N-1) \textbf{I}_N \}^{-1}$ 
			and
			$\textbf{w}^\mathrm{a}=\widetilde{\textbf{P}^\mathrm{a}}(\textbf{HX}^\mathrm{f})^\mathrm{T}\textbf{R}^{-1}(\textbf{y}-\mathcal{H}_k(\bar{\textbf{x}}^\mathrm{f}))$, and the analysis ensemble $\textbf{E}^\mathrm{a}$ is given by a linear combination of the forecast ensemble mean $\bar{\textbf{x}}^\mathrm{f}\in{\mathrm R}^{M}$ and the forecast ensemble anomalies in $\textbf{X}^\mathrm{f}$.
			Finally, $\textbf{1}^\mathrm{T}=[1,...,1]\in{\mathrm R}^{M}$ and $\mathbf{I}_N$ is the $N\times N$ identity matrix.

			The domain localization applied to the ETKF gives the {\correction LETKF of \cite{hunt07}.} 
			In the LETKF, a separate analysis is performed for each {\correction model variable attached to a gridpoint} $s\in\Gamma$ of the spatial grid and the resulting update affects each physical variable $l\in\Lambda$ at $s$.  
			The DA is performed using localized observations, i.e., the observation precision matrix is $\widetilde{\textbf{R}}_{|s}^{-1}$. 
			
			\subsection{Domain-localized CME} 
			\label{LocaLikeli}
			
			In order to formulate the CME with domain localization, we start with the general case of an observation subvector of generic nature. In section~\ref{CMEsubvector}, the case of the trivial evidencing window $K=1$ is treated first. 
			Second, we address the inconsistency that arises when $K \ge 2$ and show
			how to remove it by using realistic approximations.  
			In section~\ref{CMELocal}, we derive from the general case a formulation of the local CME. 
			Finally, in section~\ref{CMEDLCME}, we propose a heuristic global {\correction model version selection} indicator based on local CMEs, the DL-CME. 
			
			\subsubsection{CME of an observation subvector}
			\label{CMEsubvector}
			
			Let us consider first the problem of estimating the CME of $\widehat{\mathbf{y}}_{K:1}$, a subvector of the original
			observation vector ${\mathbf{y}}_{K:1}$.  
			Such a subvector may be obtained either by selecting a subset of observation types or by a spatial localization or both.    
			The corresponding CME is just a particular case of the original DA, i.e., of the DA for the complete observation vector,
			 {\correction and can be obtained by directly applying Eq.~\eqref{eq_CMEorig} to the subvector $\widehat{\mathbf{y}}_{K:1}$.}
			Indeed, if the evidencing window has length $K=1$, the contextual evidence of
			$\,\widehat{\mathbf{y}}_{1}$ becomes
			\begin{equation}
				p(\widehat{\mathbf{y}}_1|\mathbf{y}_{0:})=\int \! \mathrm{d}\mathbf{x}_0\,\, p(\widehat{\mathbf{y}}_1|\mathbf{x}_0)p(\mathbf{x}_0|\mathbf{y}_{0:}).
			\end{equation}
			Both pdfs in the integrand above are by-products of the original DA:
			$p(\mathbf{x}_0|\mathbf{y}_{0})$ is the original posterior pdf at time $t_0$ and the sub-sampled likelihood
			$p(\widehat{\mathbf{y}}_1|\mathbf{x}_0)$ also stems from the original DA process.  
			Using the EnKF formalism, one can compute the CME for this subvector using Eq.~(\ref{EnKFLikelihoodFormulation}) which in this case leads to 
			\begin{align}
				\begin{split}  
					p(\widehat{\mathbf{y}}_{1}|\mathbf{y}_{0:})&\approx (2\pi)^{-\frac{d}{2}}|\widehat{\boldsymbol{\Sigma}}_1|^{-\frac{1}{2}} \\ 
					& \! \! \! \! \! \! \! \! \! \! \! \!  \times \exp\left\{-\frac{1}{2}[\widehat{\mathbf{y}}_1-\widehat{\mathcal{H}}_1(\mathbf{x}^\mathrm{f}_1)]^\mathrm{T}
					\widehat{\boldsymbol{\Sigma}}_1^{-1}[\widehat{\mathbf{y}}_1-\widehat{\mathcal{H}}_1(\mathbf{x}^\mathrm{f}_1)]\right\}.
				\end{split}
				\label{CME_K1}
			\end{align}  
			Here $\widehat{\boldsymbol{\Sigma}}_1=\widehat{\mathbf{H}}_1\mathbf{P}^\mathrm{f}_1\widehat{\mathbf{H}}_1^\mathrm{T}+\widehat{\mathbf{R}}$,
			where $\widehat{\mathcal{H}}_1$, $\widehat{\mathbf{H}}_1$ and $\widehat{\mathbf{R}}$ are the observation operator, its
			linearization and the observation error covariance matrix corresponding to the observation subvector, respectively.  
			Note that $(\mathbf{x}^\mathrm{f}_1,\mathbf{P}^\mathrm{f}_1)$ are the forecast products of the original DA. 
			
			If the evidencing window is longer, $K>1$, we decompose the contextual evidence -- {\correction according to \cite{carrassi17}} -- as 
			\begin{align}
				\begin{split}
					p(\widehat{\mathbf{y}}_{K:1}|\mathbf{y}_{0:})  = & p(\widehat{\mathbf{y}}_{1}|\mathbf{y}_{0:}) \prod_{k=2}^{K} p(\widehat{\mathbf{y}}_{k}| \widehat{\mathbf{y}}_{k-1:0}, \mathbf{y}_{0:}) \\
					= &\int \! \mathrm{d}\mathbf{x}_1\,\, p(\widehat{\mathbf{y}}_1|\mathbf{x}_0)p(\mathbf{x}_0|\mathbf{y}_{0:}) \\
					&\times \prod_{k=2}^{K}  \int \! \mathrm{d} \mathbf{x}_{k}\,\, p(\widehat{\mathbf{y}}_k|\mathbf{x}_{k-1})p(\mathbf{x}_{k-1}|\widehat{\mathbf{y}}_{k-1}) .
				\end{split}
				\label{CMEtilde}
			\end{align}
			The $p(\widehat{\mathbf{y}}_{1}|\mathbf{y}_{0:})$ factor in Eq.~(\ref{CMEtilde}) corresponds to the case $K=1$ and it can be computed from the original DA process using Eq. (\ref{CME_K1}).  
			The sub-sampled likelihoods $p(\widehat{\mathbf{y}}_k|\mathbf{x}_{k-1})$
			are also easily computable since the likelihood of the observation is usually given in a state-estimation process.  
			
			The pdf of the system's state, conditional on the observation subvectors $p(\mathbf{x}_{k-1}|\widehat{\mathbf{y}}_{k-1})$ for $2\le k
			\le K$, however, requires a dedicated DA of the observation subvectors. 
			It seems desirable to estimate these exact pdfs but the forecast-assimilation process required to do so may --- depending on the nature of the observation subvectors --- involve an insufficiently observed system and the DA algorithm
			may, therefore, fail to converge (cf. section~\ref{DAimplSPEEDY}). 
			   
			In such a situation, approximating 
			$p(\mathbf{x}_{k-1}|\widehat{\mathbf{y}}_{k-1})$ by the original posterior pdf,
			$p(\mathbf{x}_{k-1}|{\mathbf{y}}_{k-1})$, is a possibility when the evidencing window $K$ is short enough.  
			In this case, the EnKF-formulation can be used to compute the CME as 
			\begin{align}
				\begin{split}  
					p& (\widehat{\mathbf{y}}_{K:1}|\mathbf{y}_{0:})\simeq \prod_{k=0}^K (2\pi)^{-\frac{d}{2}}|\widehat{\boldsymbol{\Sigma}}_k|^{-\frac{1}{2}} \\ 
					&\times \exp\left\{-\frac{1}{2}[\widehat{\mathbf{y}}_k-\widehat{\mathcal{H}}_k(\mathbf{x}^\mathrm{f}_k)]^\mathrm{T}
					\widehat{\boldsymbol{\Sigma}}_k^{-1}[\widehat{\mathbf{y}}_k-\widehat{\mathcal{H}}_k(\mathbf{x}^\mathrm{f}_k)]\right\}, 
					\label{CMEtilde2}
				\end{split}
			\end{align}  
			where
			$\widehat{\boldsymbol{\Sigma}}_k=\widehat{\mathbf{H}}_k\mathbf{P}^\mathrm{f}_k\widehat{\mathbf{H}}_k^\mathrm{T}+\widehat{\mathbf{R}}$.
			Using instead of the above approximation the correct pdfs of the system's state, conditional on the observation subvector $p(\mathbf{x}_{k-1}|\widehat{\mathbf{y}}_{k-1})$, would require computing the ad-hoc
			$\widehat{\mathbf{P}}^\mathrm{f}_k$ in $\widehat{\boldsymbol{\Sigma}}_k$, instead of $\mathbf{P}^\mathrm{f}_k$.
			
			\subsubsection{Local CME}
			\label{CMELocal}
			
			In the same spirit as the local analysis above, we want to locally compute the CME at each spatial gridpoint $s\in\Gamma$. 
			The reduced observation vectors $\mathbf{y}_{k|s}$ can be considered as observation subvectors $\widehat{\mathbf{y}}_k$ and we can use, therewith, the formalism of section~\ref{CMEsubvector}.
			
			Hence, for each spatial gridpoint $s$, the local CME can be obtained from Eq. (\ref{CMEtilde2}), yielding
			\begin{align}
				\begin{split}
					p&(\mathbf{y}_{K:1|s}|\mathbf{y}_{0:}) \approx\prod_{k=0}^K (2\pi)^{-\frac{d}{2}}|\widetilde{\boldsymbol{\Sigma}}_k|^{-\frac{1}{2}}\times\\& \exp\left\{-\frac{1}{2}({\mathbf{y}_k}_{|s}-\mathcal{H}_k(\mathbf{x}^\mathrm{f}_k)_{|s})^\mathrm{T}
					\widetilde{\boldsymbol{\Sigma}}_k^{-1}({\mathbf{y}_k}_{|s}-\mathcal{H}_k(\mathbf{x}^\mathrm{f}_k)_{|s})\right\};
					\label{DomainLikeliEq}
				\end{split}
			\end{align}  
			here
			$\widetilde{\boldsymbol{\Sigma}}_k={\mathbf{H}_k}_{|s}\mathbf{P}^\mathrm{f}_k{\mathbf{H}_k^\mathrm{T}}_{|s}+\widetilde{\mathbf{R}}_{|s}$;  $\mathcal{H}_k(\cdot)_{|s}$ and ${\mathbf{H}_k}_{|s}$ are, respectively, the restrictions of the nonlinear and the linearized observation operator to the neighborhood
			of $s$.  
			In other words, at the spatial point $s$, the local CME returns the CME of the observations that have affected its
			analysis.

			{\correction
				{\correction Analogously} to the LETKF, this approach computes a local CME at each gridpoint. 
				Another approach would be to taper directly the global matrix $\boldsymbol{\Sigma}_k$ {\correction -- as in the background covariance localization of Sec. \ref{CovLoc} below.}
				However, the high dimension of the systems {\correction we are interested in may not allow one} to compute its inverse, $\boldsymbol{\Sigma}_k^{-1}$. 
				The local CME approach may seem costly but, in practice, the numerical gain of computing the CME on a small number of observations, within a local domain, overcomes the great number of local computations needed. 
				Moreover, the local computation of the CME allows a full parallelization {\correction of the algorithm,} which offers an important computational gain.
			}
			
			Note that, in the case of domain localization, it is not necessary to approximate the conditional pdf of the system's state on the observation subvectors by the original
			posterior pdf, since ${\mathbf{H}_k}_{|s}\mathbf{P}^\mathrm{f}_k{\mathbf{H}_k^\mathrm{T}}_{|s}$ is already the local approximation of
			$\mathbf{H}_k\mathbf{P}^\mathrm{f}_k\mathbf{H}_k^\mathrm{T}$.  
			Nonetheless, the $\mathbf{y}_{k|s}$ observation vectors represent a \emph{tube} of observations spread in time over the evidencing window but confined to the same local domain. 
			Due, however, to the physical evolution of the error covariances, such a static tube may not be as {\correction useful for DA and evidencing as would} a dynamic tube that follows local domains {\correction and is} covariant with the flow \citep{bocquet16}.  
			{\correction
				Nevertheless, the implementation of such covariant localization techniques is not required
			 for the short evidencing windows used here,} 
			so that we apply Eq.~\eqref{DomainLikeliEq} without such
			corrections.
			
			\subsubsection{The DL-CME}
			\label{CMEDLCME}
			Creating a global indicator from local CMEs is not theoretically straightforward.  
			We propose instead a global heuristic indicator for {\correction model version selection} $\widetilde{p}(\mathbf{y}_{K:1}|\mathbf{y}_{0:})$, that we call the DL-CME,
			\begin{equation}
				\widetilde{p}(\mathbf{y}_{K:1}|\mathbf{y}_{0:})=\exp\left\{\sum_{s\in \Gamma} w(s)  \ln\{p(\mathbf{y}_{K:1|s}|\mathbf{y}_{0:})\} \right\};
			\end{equation} 
			here $w(s)$ are positive, model grid--dependent coefficients, inversely proportional to {\correction the number of observations in the domain around $s$}, that weight the logarithms of the local CMEs, such that $\sum_{s\in\Gamma}w(s)=1$. 
			For instance, in the L95 experiments of section~\ref{NumExpL95}, all the $w(s)$ weights are set to be equal since the localization radius is constant throughout the domain. 
			However, in the SPEEDY experiments (section~\ref{NumExpSPEEDY}), $w(s)$ vary with the latitude as the number of gridpoints per local domain does.  
			
		\subsection{Brief review of {\correction background} covariance localization.}
		\label{CovLoc}
		
		{\correction Another widely used technique for localization is the background covariance localization originally proposed by \cite{houtekamer01}. }
		This localization consists in restricting the spatial impact of the unrealistic correlations due to the subsampling of the ensemble-based covariances, by directly applying a smoothing function onto the forecast error covariance matrix. 
		 
		This technique can be applied to the computation of the CME in the same vein as DL-CME above, in order to take into account the actual covariances used by the DA. 
		Details of the derivation of the CME formulation using {\correction background} covariance localization are presented in \nameref{Appendix2}.
		The computational cost remains the same as for the original {EnKF-formulation} plus the significant cost of a $M\times M$ Schur product. 
		
		This alternative approach to CME localization can also be applied to large-dimensional systems, but only if combined with sophisticated numerical techniques to make it computationally feasible.
		For this reason, {\correction background} covariance localization will not be discussed furthermore in the present article. 
		Further studies investigating this matter should follow. 
		See \citealp{carrassi18} (their section~4.4) for a review on localization techinques in DA. 
		
		
	\section{Numerical experiments with a toy model} 	
	\label{NumExpL95}
The numerical experiments are designed to: (i) study numerically the performance of the global and local CME as a {\correction model version selection} indicator and to compare it with the RMSE that is widely used for the same purposes; and (ii) analyze the impact of domain localization in the computation of the CME. 
These experiments use the L95 model \citep{lorenz95,lorenz98}.
	
		\subsection{Experimental setup}
			\subsubsection{The Lorenz-95 (L95) model}
			The one-dimensional L95 model is {\correction a 
			representation} of atmospheric flow along a mid-latitude zonal circle with $M=40$ variables $\{x_i : i=1,...,40\}$. 
			In this case, $\mathrm{Card}(\Gamma)=M$ and $\mathrm{Card}(\Lambda)=1$, i.e., one single physical variable, such as {\correction the stream function or the geopotential height,}
			is defined at each grid point. 
			
			The model equations, according to \cite{lorenz98}, are the following
							\begin{equation}
							\frac{\mathrm{d}x_i}{\mathrm{d}t}=(x_{i+1}-x_{i-2})x_{i-1}-x_i+F,
							\label{Lorenz95eq}
							\end{equation}
							for $i=1,...,M$ and F represents the external forcing. 
							The model integration is performed using a fourth-order Runge-Kutta scheme with a time step $\delta t=0.05$, in nondimensional units for which unity corresponds to 5 days \citep{lorenz95}. 
			
			In our {\correction model version selection} problem, we consider the correct {\correction model version,} given by Eq.~(\ref{Lorenz95eq}) with $F=F_1=8$. 
			This value leads to chaotic behavior of the model, with a doubling time of errors equal to 2.1 days \citep{lorenz95}. We then generate various incorrect {\correction model versions}, each of them with a different forcing term $F_0\neq F_1$. 
			
			\subsubsection{The DA implementation}
			\label{L95_DAsetup}
							{\correction A ``true" trajectory is generated with the correct} {\correction model version, in which $F=F_1=8$.} 
							Observations are generated by sampling the true trajectory $\textbf{x}^\mathrm{t}$, every DA interval $\Delta t = 0.05$, using Eq. (\ref{Eq_SetofObs}) with $\epsilon\sim\mathcal{N}(0,1)$ and $\textbf{H}_k=\textbf{I}_{40}$, so that the full system is observed. The correct observation error covariance, $\textbf{R}=\textbf{I}_{40}$, is used in the DA.
							The experiments are run over $T=5\times10^4$ DA cycles, after a spin-up of $T_{{\rm su}}=10^4$ DA cycles. 
							
							The DA method used throughout the present section is the LETKF described in section \ref{LETKFmethodo}. 
							The localization matrix, $\bf{L}$, is diagonal with coefficients defined by $\left\{G\left(|\widetilde{d}/2-i|/\rho_\mathrm{loc}\right)\right\}_{i,i}$, where $\widetilde{d}$ is the size of the reduced observation error covariance matrix, $\rho_\mathrm{loc}$ is the localization radius and $G$ is the Gaspari-Cohn function \citep{gaspari99} displayed in Eq. (\ref{GaspariAndCohn}) of \nameref{Appendix2}. 
							 		
							The performance of the DA algorithm is evaluated using the RMSE of the analysis with respect to the truth
							\begin{equation}
							\text{RMSE}^\mathrm{t} = \Bigg\{ \frac{1}{M}\sum_{i=1}^{M} \big(x^\mathrm{a}_i-x^\mathrm{t}_i \big)^2 \Bigg\}^{1/2},
							\label{RMSEt}
							\end{equation}
							with $\textbf{x}^\mathrm{t} = (x^\mathrm{t}_i)_{i=1:M}$ and $\textbf{x}^\mathrm{a} = (x^\mathrm{a}_i)_{i=1:M}$ being the true and  {\correction the estimated state, respectively.} Here and elsewhere we follow the superscript notation $\{{\rm t, a, f, o}\}$ proposed by \cite{ide97} for the truth; the analysis, i.e. the estimated state; the forecast, {\correction i.e. the propagated state between consecutive observations, with $\Delta t = 0.05$; and} the observations.
							
							The accuracy of the {\correction model version selection} is assessed based on the RMSE of the forecast with respect to the observations
							\begin{equation}
							\text{RMSE} = \Bigg\{ \frac{1}{d}\sum_{i=1}^{d} \big(\mathcal{H}(x^\mathrm{f}_i)-y_i \big)^2 \Bigg\}^{1/2}.
							\label{RMSE}
							\end{equation} 
							   
							Model evidence, for both the G-CME and the DL-CME formulations, is evaluated over the evidencing window {\correction $\{t_1,...,t_K\}$ of size $K$}.
							Unless otherwise stated, we set here  $K=1$, which means that the model evidence is computed only for one DA cycle. 
							{\correction
							This choice is made based on the fact that, in a selection problem, the smaller the evidencing window the more difficult it is to discriminate between two model {\correction versions} \citep{carrassi17}.                                         
							In practice, this means that the assimilation process is performed repeatedly in time and, at each available observation, the discrimination skills of the selection methods are assessed at the present time step, i.e., for the most difficult {\correction (i.e. least discriminating) selection problem, namely $K=1$.}
						}
							
			\subsection{Numerical results}
			
				\subsubsection{CME versus RMSE}
				In this subsection, we focus on the performance of the G-CME (Eq.~\ref{EnKFLikelihoodFormulation}) --- i.e., without localization --- as the {\correction model version selection} indicator, and compare it with that of the RMSE. 
				
				For each {\correction model version}, i.e. for each choice of the forcing F, a DA is performed using the setup described in section \ref{L95_DAsetup}, with a 40-member ensemble ($N=40$), without localization and with a coefficient of inflation tuned to provide the smallest RMSE$^\mathrm{t}$ (Eq.~\ref{RMSEt}) for each DA.
				At each cycle of the experiment, the G-CME and RMSE indicators are computed.

				\begin{center}
					\begin{figure}[h!]
						\centering
						\includegraphics[width=8cm]{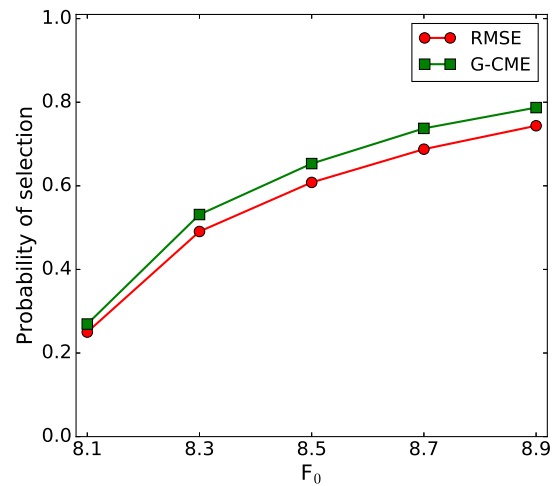}  
						\caption{Probability of selecting the correct {\correction model version}, with $F_1=8$, vs. a particular incorrect {\correction model version},
							with $F_0=8.1, 8.3, 8.5, 8.7$ 
							and $8.9$. The probabilities are calculated using one of the two {\correction model version selection} indicators: 
						RMSE (solid red line and circles) and G-CME (solid green line and squares),
						for a 40-member ensemble.
						} 
						\label{Figure1}
					\end{figure} 
				\end{center}

				\paragraph{Probability of selection.} 
								 
				We first look at the probability of selection of the correct {\correction model version} for each indicator; that is, $2R-1$ where $R$ is computed by counting how many times each indicator chooses the correct {\correction model version} over the incorrect one, out of the total 
				of $T=5\times10^4$ cycles. 
				Thus, a random classifier will produce a probability of selection equal to $0$ and a perfect classifier will produce one equal to $1$. 
				This allows us to evaluate the reliability of the indicators in terms of {\correction model version selection}.  
				Then, by repeating the computation for different values of $F_0$, we deduce the sensitivity of either indicator to the discrepancy $F_1$ - $F_0$ between correct and incorrect {\correction model version}.

				The probabilities of selection are given in Figure \ref{Figure1}. 
				The incorrect {\correction model versions} are simulated with the forcings $F_0=8.1,8.3,8.5,8.7,8.9$ plotted on the $x$-axis. 
				
				Note that the larger the forcing $F_0$, the easier it should be to select the correct {\correction model version} over the incorrect one. 
				Therefore, a good {\correction model version selection} indicator is the one that increases the fastest when $F_0$ is increased. 
				
				From Figure~\ref{Figure1}, we see that the RMSE selects the correct {\correction model version} over the incorrect one given by $F_0=8.1$ with a probability of $0.25$ that is slightly inferior to the G-CME probability of $0.27$.  
				As the forcing of the incorrect {\correction model version} increases, both indicators increase their probabilities of selection but the growth of the G-CME indicator is faster.  
				  
				 \begin{center}
				 	\begin{figure*}[t!]
				 		\centering
				 		\begin{tabular}{cc}
				 			\includegraphics[height=7cm,width=8cm]{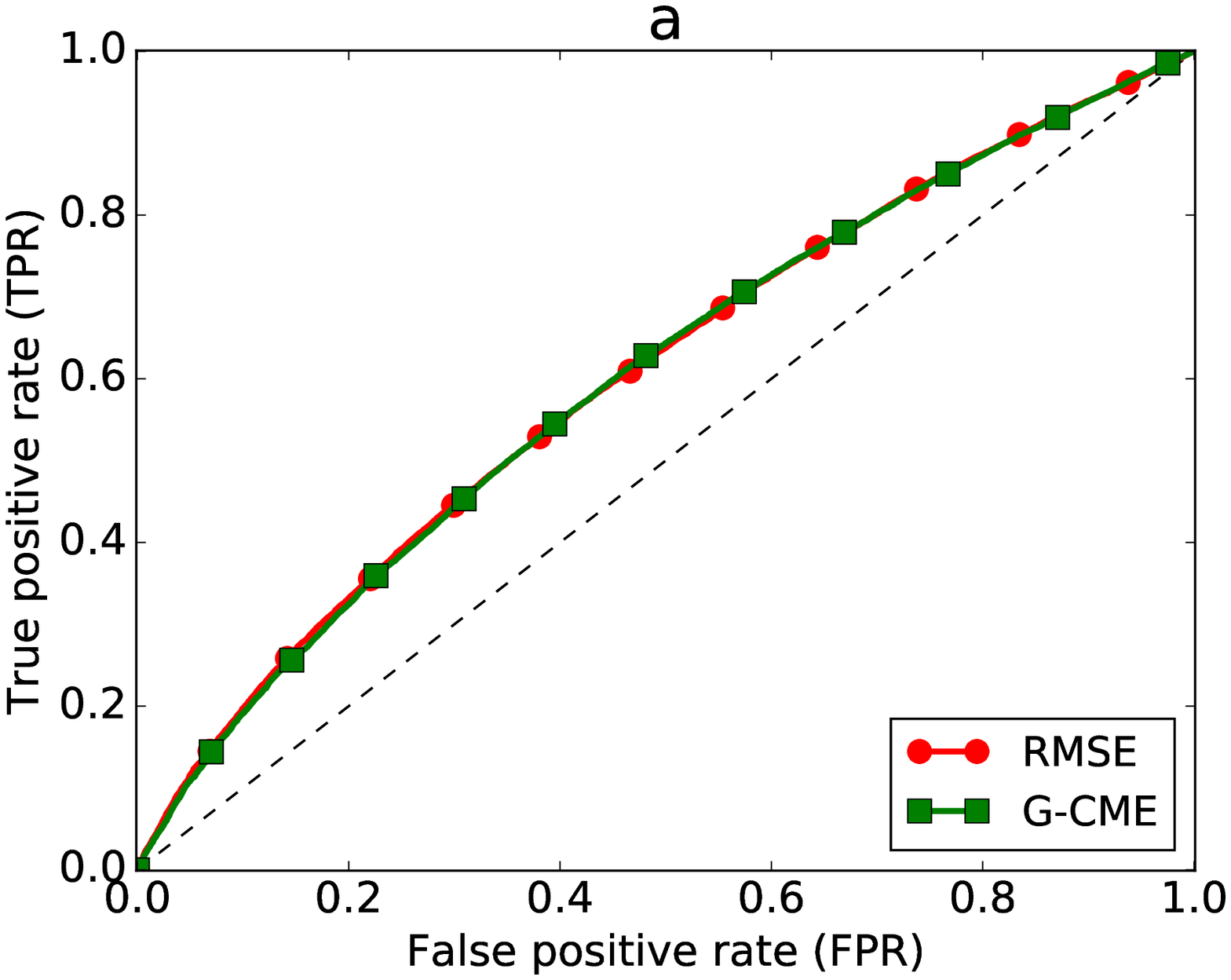} &
				 			\includegraphics[height=7cm,width=8cm]{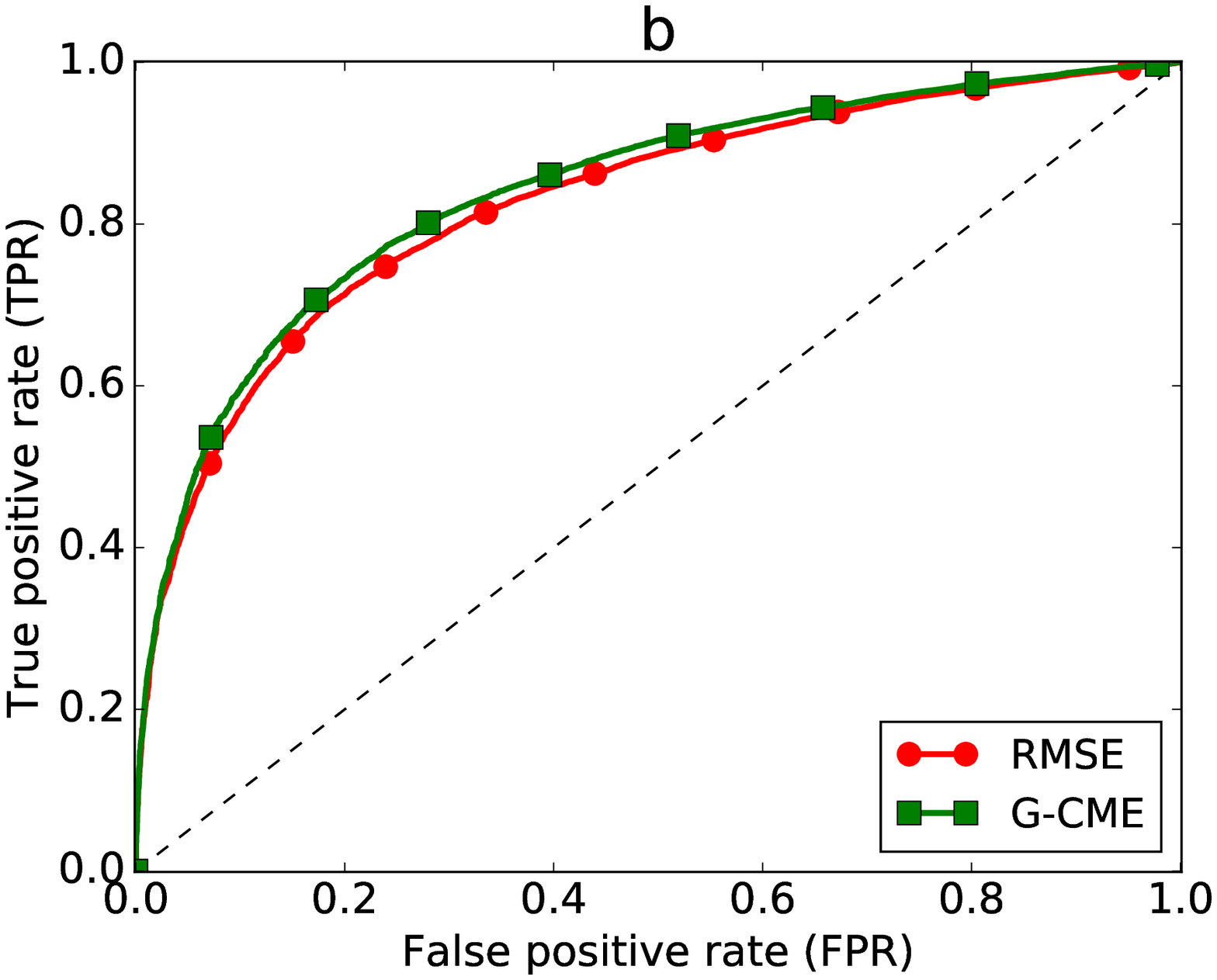}  
				 		\end{tabular} 
				 		\caption{
				 			ROC curves of the two {\correction model version selection} indicators, RMSE and G-CME between the correct {\correction model version}, with $F_1=8$, and two of the incorrect {\correction model versions}, with (a) $F_0=8.1$, and (b) $F_0=8.9$, both for a 40-member ensemble. 
				 			The two curves, for RMSE and G-CME, use the same lines as in Figure~\ref{Figure1}.
				 			}
				 		\label{Figure2}
				 		
				 	\end{figure*} 
				 \end{center} 		
				
				\paragraph{ROC curves and the Gini coefficient.} 
				The probabilities of selection as computed in Figure \ref{Figure1} only evaluate the ability of the indicators to select one {\correction model version} over the other. 
				Yet the values of these indicators also provide distances between the two {\correction model versions.} 
				The greater these distances, the greater the confidence in the corresponding {\correction model version selection}.  

				We define the confidence value in the {\correction model version selection} for the G-CME indicator as the difference between the logarithms of the model evidences, introduced in section \ref{ModEvidAndDA}, namely
				\begin{equation}
				\Delta_\text{G-CME}=\ln\{p_1(\textbf{y})\}-\ln\{p_0(\textbf{y})\}, 
				\end{equation}
				with $p_1$ being the G-CME value for the correct {\correction model version} and $p_0$ the G-CME value for the incorrect one.   
				Similarly, a confidence value is defined for the RMSE, as
				\begin{equation}
				 \Delta_\text{RMSE}=\text{RMSE}_0-\text{RMSE}_1, 
				 \end{equation}
				 with RMSE$_1$, defined in Eq. (\ref{RMSE}), for the correct {\correction model version} and RMSE$_0$ for the incorrect one.	
				  
				In order to evaluate the ability of each indicator to select the correct {\correction model version} with various levels of confidence, we compute a {\correction Receiver (or Relative)} Operating Characteristic (ROC) curve. 
				A ROC curve measures the performance of a binary classifier; see, for instance, \cite{metz78} for a review of ROC analysis.  

				In our study, we define the ROC curve as follows.
				If the confidence value $\Delta$ (either $\Delta_\text{G-CME}$ or $\Delta_\text{RMSE}$) is positive, we consider the correct {\correction model version} to have been selected and the incorrect one if it is negative. 
				Knowing that there is only one correct {\correction model version}, we define the selection of the latter as a {true positive} instance and the selection of the incorrect {\correction model version} as a {false positive} instance. 
				
				Note that, with this definition of the classifier, there is neither a true negative nor a false negative for which a confidence value is obtained. 
				Still, this configuration allows us to define the {True Positive Rate} (TPR) and the {False Positive Rate} (FPR), two parametric functions of the threshold $\Xi\in[0,1]$, giving the proportion of positives  that are correctly identified as such (i.e., when $\Delta>\Xi$) and, respectively, the proportion of negatives wrongly identified as positives (i.e., when $-\Delta>\Xi$).   
				The ROC curve is then displayed by plotting parametrically the TPR($\Xi$) against the FPR($\Xi$)  for varying thresholds $\Xi$. 
				
				The random classifier has a $50\%$ chance of selection of each {\correction model version} for every threshold and it will produce a diagonal ROC curve (black dashed line in Figures \ref{Figure2}, \ref{Figure4} and \ref{Figure6}), namely a straight line between the origin $(0,0)$ and the point $(1,1)$, while a perfect classifier that always selects the correct {\correction model version has} a ROC curve equal to $0$ for $\Xi =0$ and to $1$ for $\Xi \in (0,1]$. 
				
				Figure \ref{Figure2} shows the ROC curves of the {\correction model version selection} indicator RMSE (solid red line with red circles) and the {\correction model version selection} indicator G-CME (solid green line with squares), for two incorrect {\correction model versions}: (a) the one closest to the correct value, i.e. $F_0=8.1$; and (b) the one farthest from it, i.e. $F_0=8.9$.
				Both indicators outperform random selection by exhibiting a higher TPR/FPR ratio throughout, for every threshold $\Xi$.  
				{\correction As was the case for} the probability of selection in Figure~\ref{Figure1}, the RMSE and the G-CME are almost equally selective for the more difficult case of the incorrect {\correction model version} $F_0=8.1$ in panel (a), while the G-CME indicator slightly outperforms the RMSE for $F_0=8.9$ in panel (b).
				
				The {\correction \cite{gini21} coefficient $\gamma$} provides a simple scalar that 
				quantifies the improvement of a given classifier over the random one.  
				This coefficient equals the area $A$ between the  classifier's ROC curve and the diagonal of the graph --- which is the ROC curve of the random classifier --- divided by the area between the latter and the $x$-axis. 
				In other words, $\gamma = A/(1/2) = 2 \times A$. 
				Hence, the Gini coefficient for the random classifier is $0$ and the Gini coefficient for a perfect classifier is $1$.

				\begin{center}
					\begin{table}[h!] 
						\centering
						$
						\begin{array}{|c||c|c|}  
						\hline
						\cellcolor{black!25}& \text{RMSE} & \text{G-CME} \\
						\hline
						F_0=8.1 & 0.205 & 0.202  \\
						\hline
						F_0=8.9 & 0.652 & 0.680  \\
						\hline
						\end{array}		
						$	 
						\caption{
							Gini coefficients of the {\correction model version selection} indicators obtained using the RMSE and the G-CME between the correct {\correction model version}, with $F_1=8$, and the two incorrect {\correction model versions}, with $F_0=8.1$ and $F_0=8.9$, computed for a 40-member ensemble. 
							These Gini coefficients are based on the ROC curves in Figure~\ref{Figure2}.
							}
						
						\label{Table1}
					\end{table} 
				\end{center} 
				
				The Gini coefficients corresponding to the ROC curves in Figure \ref{Figure2} are reported in Table \ref{Table1}. 
				 In the case of the incorrect {\correction model version} with $F_0=8.1$, the Gini coefficients of the RMSE and the G-CME are very close, with only a $0.003$ difference in favor of the RMSE.  
				The G-CME outperforms the RMSE in the case of the incorrect {\correction model version} with $F_0=8.9$: its Gini coefficient is higher by $0.028$.

		To summarize, results so far suggest that the G-CME and the RMSE present comparable selection skills for similar {\correction model versions}, such as in the case $F_1=8$ and $F_0=8.1$, but the G-CME converges faster to a perfect classifier as $F_0$ is increased.

				\subsubsection{CME with domain localization}
				
				In the previous subsection, a 40-member ensemble was used to compute the {\correction model version selection} indicators. 
				However, in geosciences applications, such a big ensemble, with $N$ comparable to the state vector size $M$, is quite unrealistic. 
				To mimic a more realistic scenario, we use here a 10-member ensemble,  i.e. $N \ll M = 40$, in conjunction with localization. 
				\cite{bocquet17} have shown that $N=10$ is smaller than the dimension of the unstable--neutral subspace below which localization is mandatory. 
				 
				Here we use the localization radius $\rho_\mathrm{loc}=5$, as it produces the smallest RMSE$^\mathrm{t}$ for the DA process applied to the incorrect {\correction model version} (not shown here). 
				Since the localization is tuned for the incorrect {\correction model version}--DA, but left untuned for the correct {\correction model version}--DA, the analysis errors of the two DAs should be closer to each other than if the localization were tuned independently for each {\correction model version}. 
				This choice is expected to render the {\correction model version selection} task harder.
				
				We compare then the selectivity of the DL-CME to the G-CME and the RMSE. 
				Note that unless stated otherwise the same 10-member ensemble is used to compute the three {\correction model version selection} indicators: DL-CME, G-CME and RMSE. 
				As before, these three {\correction model version selection} indicators are computed at each of the $T=5\times10^4$ cycles of the numerical experiment.

						\begin{center}
							\begin{figure}[t!]
								\centering
								\includegraphics[width=8cm]{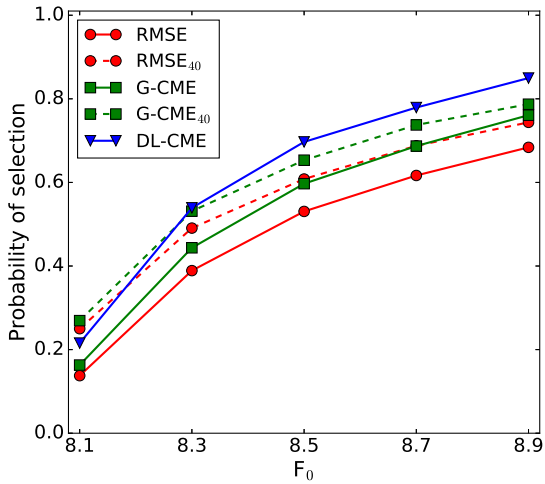}  
								\caption{
									Same as Figure~\ref{Figure1}, but including now the DL-CME {\correction model version selection} indicator, with a localization radius of $\rho_{{\rm loc}} = 5$, shown as the solid blue line with triangles. 
									All three solid lines correspond to a 10-member ensemble, while the results from Figure~\ref{Figure1} for the RMSE and the G-CME that were computed for a 40-member ensemble are plotted here as dashed; see also the figure's legend. 
								}
								\label{Figure3}
								
							\end{figure} 
						\end{center}

				\paragraph{Probability of selection.} 
				 
				  Figure~\ref{Figure3} shows the probabilities of selection as a function of the incorrect {\correction model versions}, with $F_0 > F_1$, for three {\correction model version selection} indicators: RMSE, G-CME and DL-CME. 
				  The probabilities for the RMSE and the G-CME using the 40-member ensemble, and displayed in Figure~\ref{Figure1}, are also reported in Figure \ref{Figure3} for comparison. 
				  
				  First, we note that the probabilities of selection for both RMSE and G-CME are lower for $N = 10$ than for $N = 40$. 
				  This means that the use of a smaller ensemble and no localization in the DA reduces the probabilities of a correct selection by the RMSE and the G-CME and is a direct consequence of the reduced accuracy of the DA in performing the state estimate.  
				  However, the convergence rate of the G-CME with $N = 10$ is faster than for G-CME$_{40}$ and, when $F_0=8.9$, they select the correct {\correction model version} with almost the same accuracy. 
				  
				  The key result, though, is related to the effect of localization. 
				  We see that for the same ensemble size $N=10$, the DL-CME outperforms both the RMSE and the G-CME indicators. 
				  Notably, the DL-CME with $N=10$ selects the correct {\correction model version} with a higher probability 
				  than the G-CME with $N=40$, except for the case of smallest error in {\correction model version} definition, i.e. for $F_0=8.1$.

		 		 \begin{center}
		 		 	\begin{figure*}[t!]
		 		 		\centering
		 		 		\begin{tabular}{cc}
		 		 			\includegraphics[height=7cm,width=8cm]{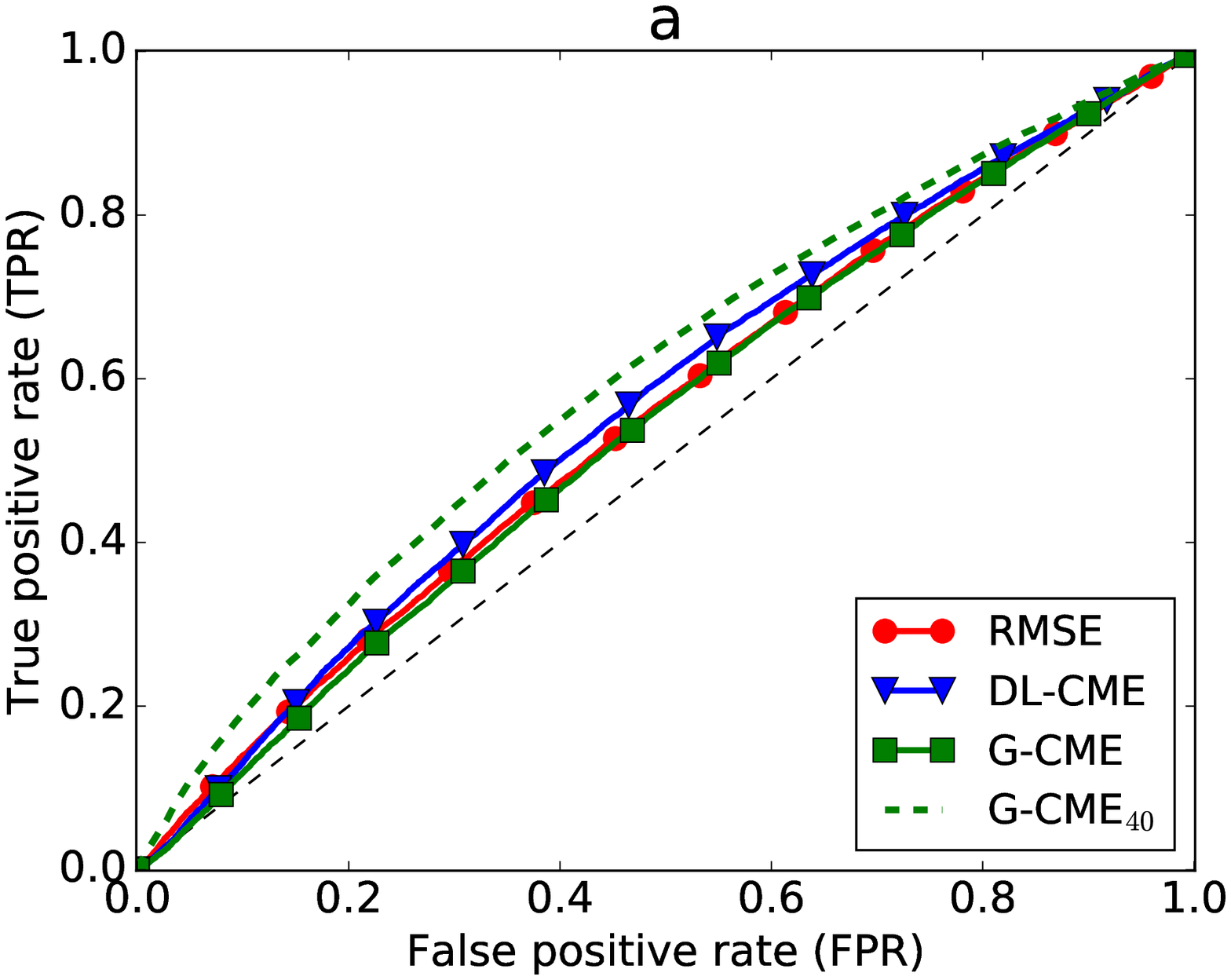} &
		 		 			\includegraphics[height=7cm,width=8cm]{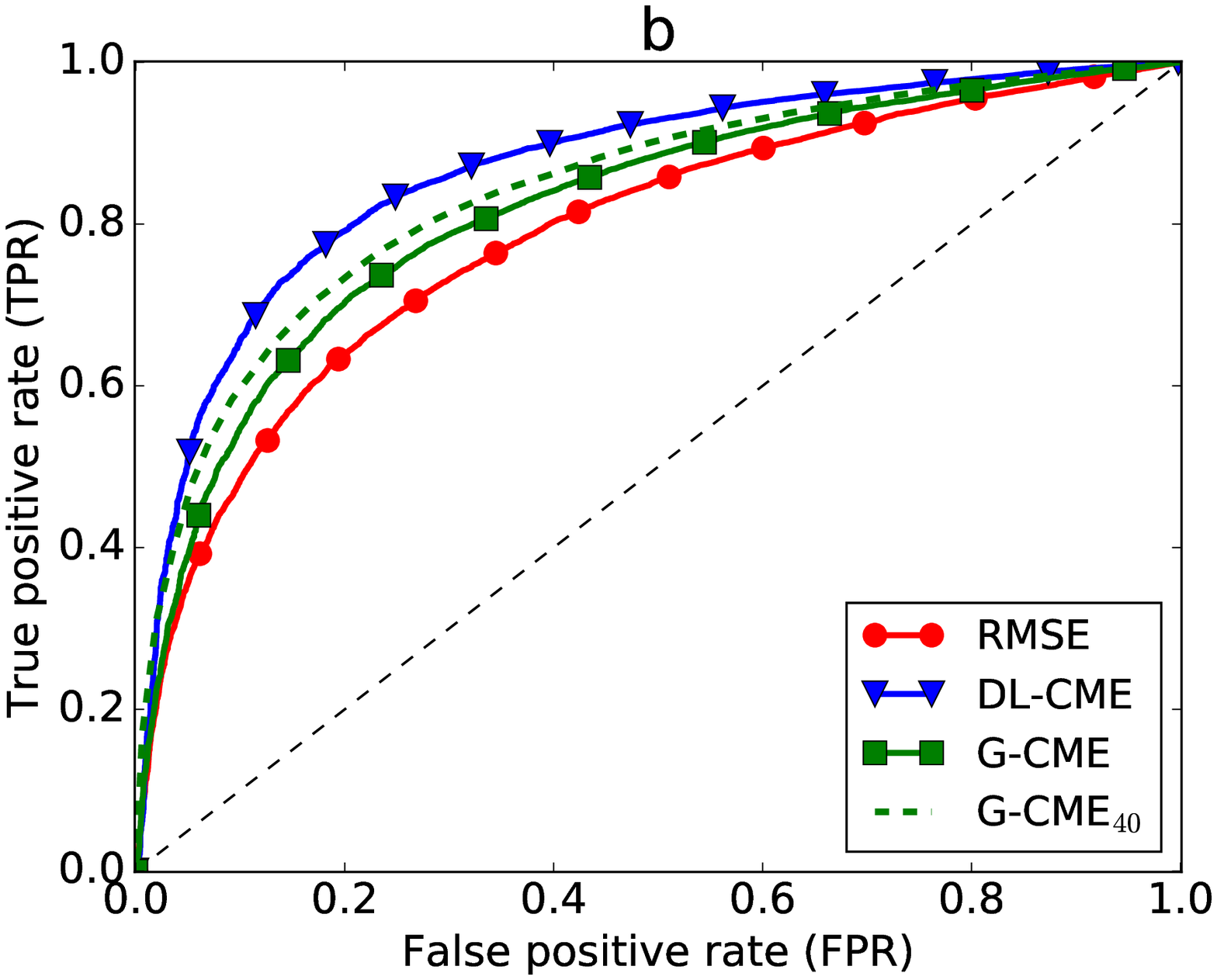} 
		 		 		\end{tabular} 
		 		 		\caption{
		 		 			ROC curves of four {\correction model version selection} indicators plotted in Figure~\ref{Figure3}. 
		 		 			The RMSE (solid red line with circles), G-CME (solid green line with squares) 
		 		 			and DL-CME (solid blue line with triangles) all use a small ensemble with $N=10$, while the G-CME$_{40}$ (dashed green line) uses a larger, 40-member ensemble. 
		 		 			As in Figure~\ref{Figure2}, the ROC curves evaluate the skill at distinguishing between the correct {\correction model version}, with $F_1=8.0$, and two of the incorrect {\correction model versions}: (a) with $F_0=8.1$, and (b) with $F_0=8.9$. 
		 		 			As in Figure~\ref{Figure2}, the DL-CME uses a localization radius of 5. 
		 		 		}
		 		 		\label{Figure4}
		 		 		
		 		 	\end{figure*} 
		 		 \end{center} 		
		 		 
		 \paragraph{ROC curves and the Gini coefficient.} 
		 The previous conclusion is confirmed by the ROC curves in the cases of the incorrect {\correction model versions} with $F_0=8.1$ and $F_0=8.9$ that are plotted, respectively, in Figures \ref{Figure4}(a, b).
		 The ROC curves of G-CME$_{40}$ have also been plotted in Figure \ref{Figure4} for reference and benchmark. 
		 The DL-CME has a TPR/FPR ratio that is larger than the G-CME for both incorrect {\correction model versions}.  
		 This result represents a substantial improvement in the computation of the CME. 
		 
		In the difficult case of $F_0=8.1$, the DL-CME with $N=10$ is not as accurate in {\correction model version selection} as the G-CME$_{40}$ but, as previously stated, the use of large ensembles is most often not feasible for models of a realistic size. 
		 Interestingly, in the case of $F_0=8.9$, the DL-CME still provides a more accurate {\correction model version selection} than the G-CME$_{40}$.
		  
		  These results are confirmed by the corresponding Gini coefficient listed in Table \ref{Table2}.   
		  In the case $F_0=8.9$, the DL-CME {\correction model version selection} skill outperforms the G-CME$_{40}$, even though the DA performance for state estimation in terms of RMSE$^\mathrm{t}$ is better when $N=40$ with no localization than $N=10$ with localization (not shown).  
		    
		    \begin{center}
		    	\begin{table}[h!] 
		    		\centering
		    		$
		    		\begin{array}{|c||c|c|c|c|}  
		    		\hline
		    		\cellcolor{black!25}& \text{RMSE} & \text{G-CME} &\text{DL-CME}& \text{G-CME}_{40}\\
		    		\hline
		    		F_0=8.1 & 0.102  & 0.097 & 0.154& 0.202 \\
		    		\hline
		    		F_0=8.9 & 0.590 & 0.656 & 0.748 & 0.680 \\
		    		\hline
		    		\end{array}		
		    		$	 
		    		\caption{ 
		    			Same as Table~\ref{Table1}, for the four ROC curves in Figure~\ref{Figure4}.
		    			}
		    		\label{Table2}
		    		
		    	\end{table} 
		    \end{center}

		 \subsubsection{Sensitivity to the localization radius}
\begin{center}
		 	\begin{figure}[h!]
		 		\centering
		 		\includegraphics[height=7cm,width=8.5cm]{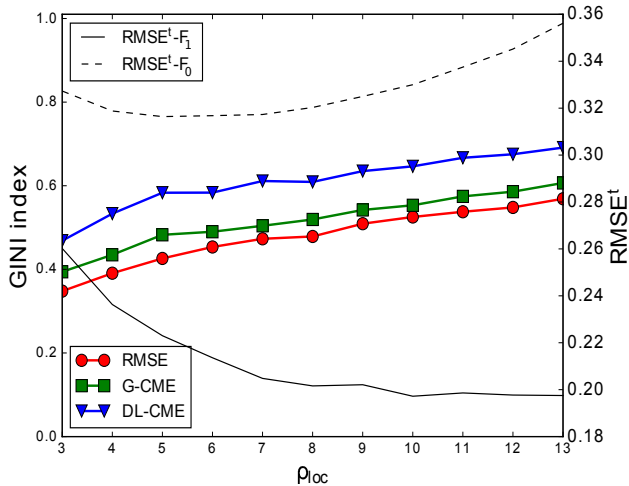}  
		 		\caption{
		 			Gini coefficients of the three {\correction model version selection} indicators between the correct {\correction model version}, with $F_1=8.0$ and the incorrect {\correction model version} with $F_0=8.5$: RMSE, G-CME, and DL-CME; each of the three indicators is plotted using the same curve styles as in Figure~\ref{Figure4}. 
		 			The results are for a 10-member ensemble and the varying localization radius $\rho_{{\rm loc}}$ is shown on the $x$-axis. 
		 			The solid black line and the dashed black line represent the RMSE$^{\rm t}$ scores for the correct and the incorrect {\correction model versions}, labeled as $F_1$ and $F_0$, respectively. 
		 		}
		 		\label{Figure5}
		 		
		 	\end{figure} 
		 \end{center}

		 Even though only the DL-CME takes into account the localization, the G-CME and the RMSE are also potentially sensitive to the localization radius since they are based on the output of a DA cycle that uses localization.  
		 The way the three {\correction model version selection} indicators respond to the localization radius $\rho_\mathrm{loc}$ is examined in Figure \ref{Figure5}.
		  
		 The experimental setup is identical to that in the previous subsections and the ensemble size is $N=10$. 
		 The localization radius $\rho_{{\rm loc}}$ is shown on the $x$-axis; while the Gini coefficient for each {\correction model version selection} indicator is plotted on the $y$-axis. 
		 Superimposed on Figure~\ref{Figure5} are the RMSE$^\mathrm{t}$ curves for the DA with the correct and the incorrect {\correction model version}, in solid and dashed black lines, respectively. 
		 
		 The results in Figure~\ref{Figure5} confirm, once more, the superiority of the two CME indicators over the RMSE. 
		 They also confirm the improvement achieved by taking into account domain localization in the CME computation, i.e., the better {\correction model version selection} by the DL-CME with respect to the G-CME; both improvements are present across all localization radii. 
		 
		 It is, moreover, worth noting that the sensitivity of all three indicators to the localization radius is small. 
		 For all the indicators, the Gini coefficient increases slowly with $\rho_{{\rm loc}}$. 
		 This increase can be explained by the difference in performance of the DA's state estimation. 
		 Indeed the difference between the two RMSE$^\mathrm{t}$ curves also increases with the localization radius.
		 Hence, the selection between the two DA-produced ensembles becomes easier as $\rho_{{\rm loc}}$ increases.
		 The key result, however, is that the sensitivities of the two CME indicators are approximately the same.
		 Thus, taking the localization into account in the CME computation should improve the CME's {\correction model version selection} skill, whatever the value of the localization radius. 
		 
		 \begin{center}
		 	\begin{figure*}[h!]
		 		\centering
		 		\begin{tabular}{cc}
		 			\includegraphics[height=7cm,width=8cm]{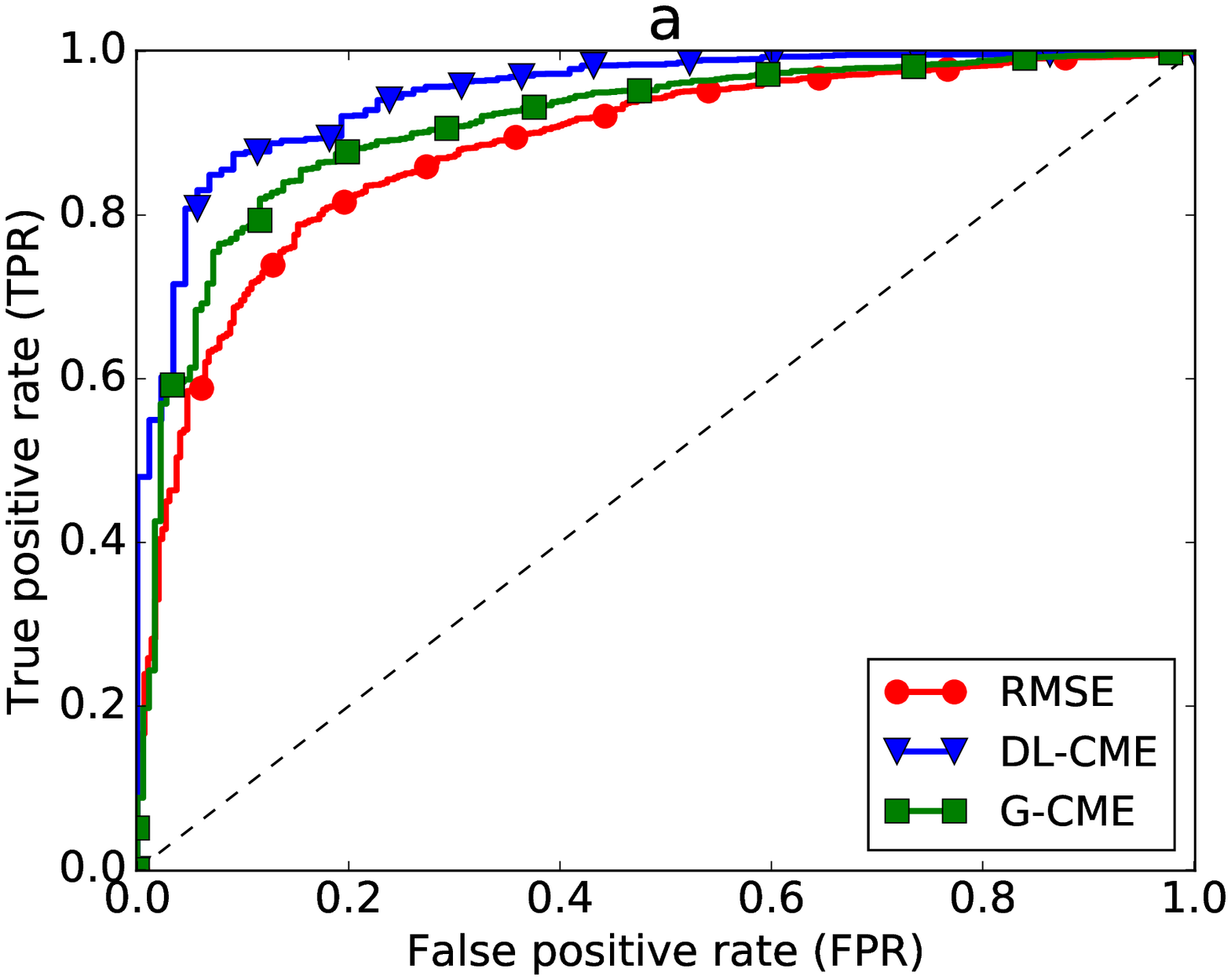} &
		 			\includegraphics[height=7cm,width=8cm]{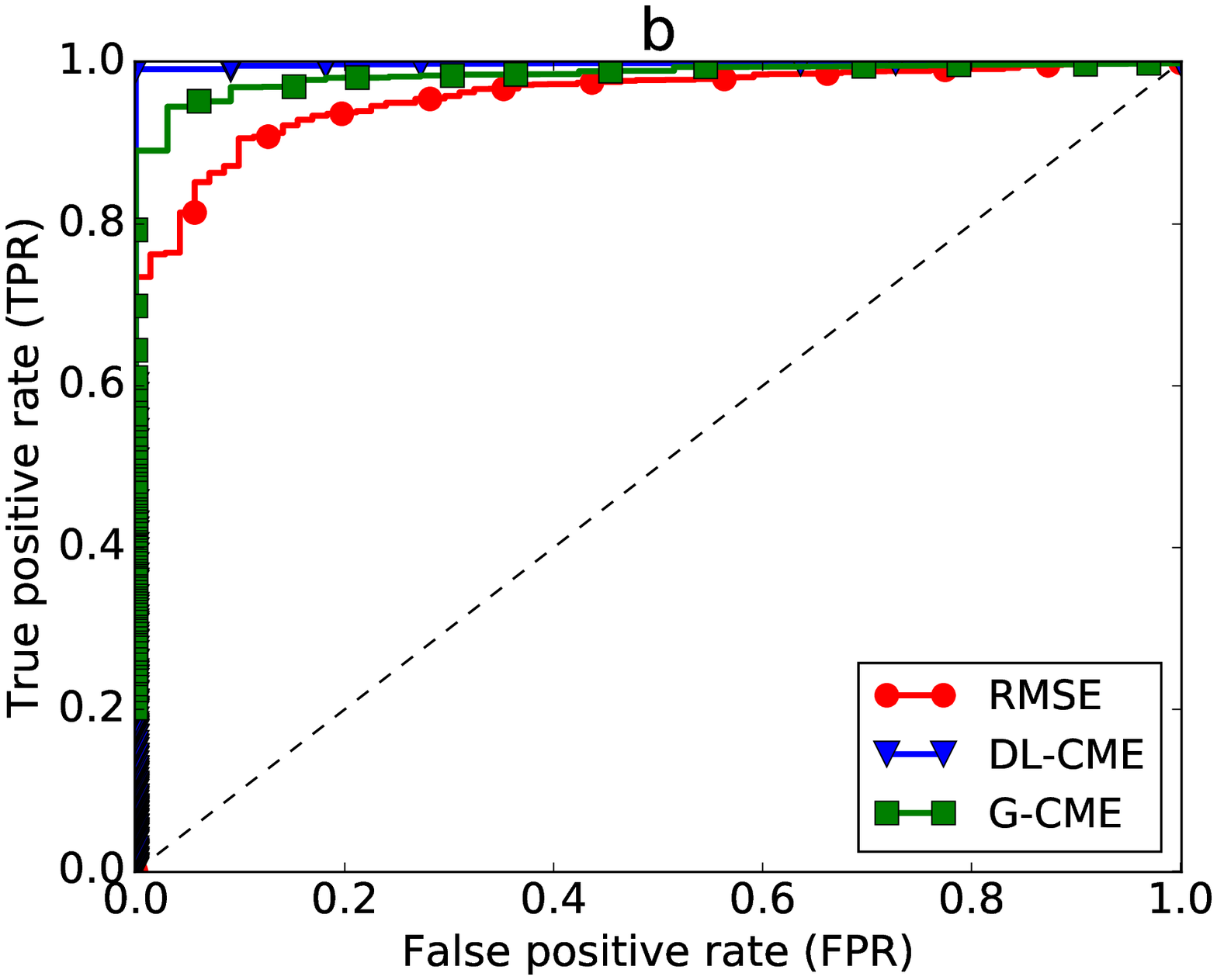} 
		 		\end{tabular} 
		 		\caption{
		 			Same as Figure~\ref{Figure4}(b), but for a longer evidencing window and using only a 10-member ensemble: (a) $K=2$, and (b) $K=4$. 
		 		}
		 		\label{Figure6}
		 		
		 	\end{figure*} 
		 \end{center} 			
		 
		 \subsubsection{Dependence on the evidencing window}
		 \label{SecEvidWind}
		 In all previous experiments, the {\correction model version selection} indicators are computed over an evidencing window $K=1$, i.e., using only one DA cycle. 
		 This choice was made to assess the indicators' performance in the most difficult setting. 
		 Indeed, selecting a {\correction model version} based on static information alone is a difficult task that is
		 not usually required in geosciences applications. 
		 In the latter, one may want to select the most accurate model representation of a meteorological event or a climatological time span, seen as a ``video loop'' or a ``full-length movie'', respectively \citep{ghil97}. 
		 We thus want now to evaluate the indicators in a more realistic setting, by comparing the correct {\correction model version}, with $F_1=8.0$, and the most incorrect one being considered, with $F_0=8.9$, over larger evidencing windows, namely $K=2$ and $K=4$. 
		 
		 The ROC curves obtained for these two experiments, with $K=2$ and $K=4$, are displayed in Figures~\ref{Figure6}(a) and (b), respectively. 
		 Comparing these curves with these for K=1 of Figure~\ref{Figure4}(b), we see that increasing the evidencing window improves the selection skills of all the indicators as they converge to the perfect classifier.  
		 This behavior is consistent with the results obtained by \cite{carrassi17} in the absence (of the need) of localization. 
		 
		 Moreover, in the case $K=2$, the DL-CME still outperforms the RMSE and the G-CME. 
		 In the case $K=4$,  both CME indicators remain more accurate than the RMSE, while the DL-CME is almost achieving a perfect selection. 
		 Altogether, the results in Figures~\ref{Figure4}(b), \ref{Figure6}(a) and \ref{Figure6}(b) indicate that when $K$ is increased all indicators converge to the perfect classifier. 
		 It follows that, for small and medium evidencing windows, the DL-CME outperforms the CME with no localization and the RMSE, but larger evidencing windows, when feasible computationally, may allow all indicators to show similar selection skills.

		
	\section{Numerical experiments with an atmospheric model} 	 
	\label{NumExpSPEEDY}
	This section presents the first implementation of the CME approach for an intermediate-complexity model. 
	The numerical experiments herein confirm the benefits of applying CME in general, and DL-CME in particular, as a {\correction model version selection} indicator for larger models.

		\subsection{Experimental setup }   
		\subsubsection{The SPEEDY model}
	We use an atmospheric general circulation model (AGCM) based on a spectral primitive-equation dynamic core that resolves the large-scale dynamics of the atmosphere. 
	The model was developed at the International Center for Theoretical Physics (ICTP) and it is referred to as the ICTP AGCM or the Simplified Parameterizations, primitivE-Equation DYnamics (SPEEDY) model (\citealp{molteni03,kucharski06}). 
	With its dynamical core and its simplified physical parameterization schemes, SPEEDY simulates successfully some of the complex physics described by state-of-the-art AGCMs, while maintaining a low computational cost \citep{neelin10}. 
	Hence, its intermediate-level complexity --- between low-order models and high-end AGCMs --- has made SPEEDY an important test bed for model development studies (e.g., \citealp{kucharski03,haarsma03,bracco04,kucharski06}) and for evaluating DA methodology (\citealp{miyoshi05, miyoshi07}). 
			   
			 The model's spectral dynamical core \citep{molteni03} uses a hydrostatic, $\sigma$-coordinate, spectral-transform formulation in the vorticity-divergence form described by \cite{bourke74}. 
			 The configuration we use here is described by \cite{miyoshi05} and it has a spectral resolution of T30L7, i.e., a horizontal spectral truncation at spherical wavenumber 30 and 7 vertical layers. 
			 It has a vertical $\sigma$-coordinate and it computes four physical variables in every vertical layer: zonal wind $u$, meridional wind $v$, temperature $T$, and relative humidity $q$, as well as one surface variable, in the lowest vertical layer, namely surface pressure $p_s$. 
			 
			 The simulated values for each vertical field are given on a 96 long. $\times$ 48 lat. horizontal grid. 
			 A summary description of SPEEDY, including its simplified physical parametrizations, is available in the appendix of \cite{molteni03}. 
			 We provide here a detailed description of its convection parametrization, since the purpose of the present section is to perform parameter selection between two values of a convective parameter.  
			 In a sense, CME and DL-CME can be viewed also in the framework of the parameter estimation problem (\citealp{ghil91}; \citealp{ghil97}; \citealp{navon97}; \citealp{carrassi17}).
			
			\begin{center}
				\begin{figure*}[h!]
					\centering 
					\begin{tabular}{cc}
						\includegraphics[width=8cm]{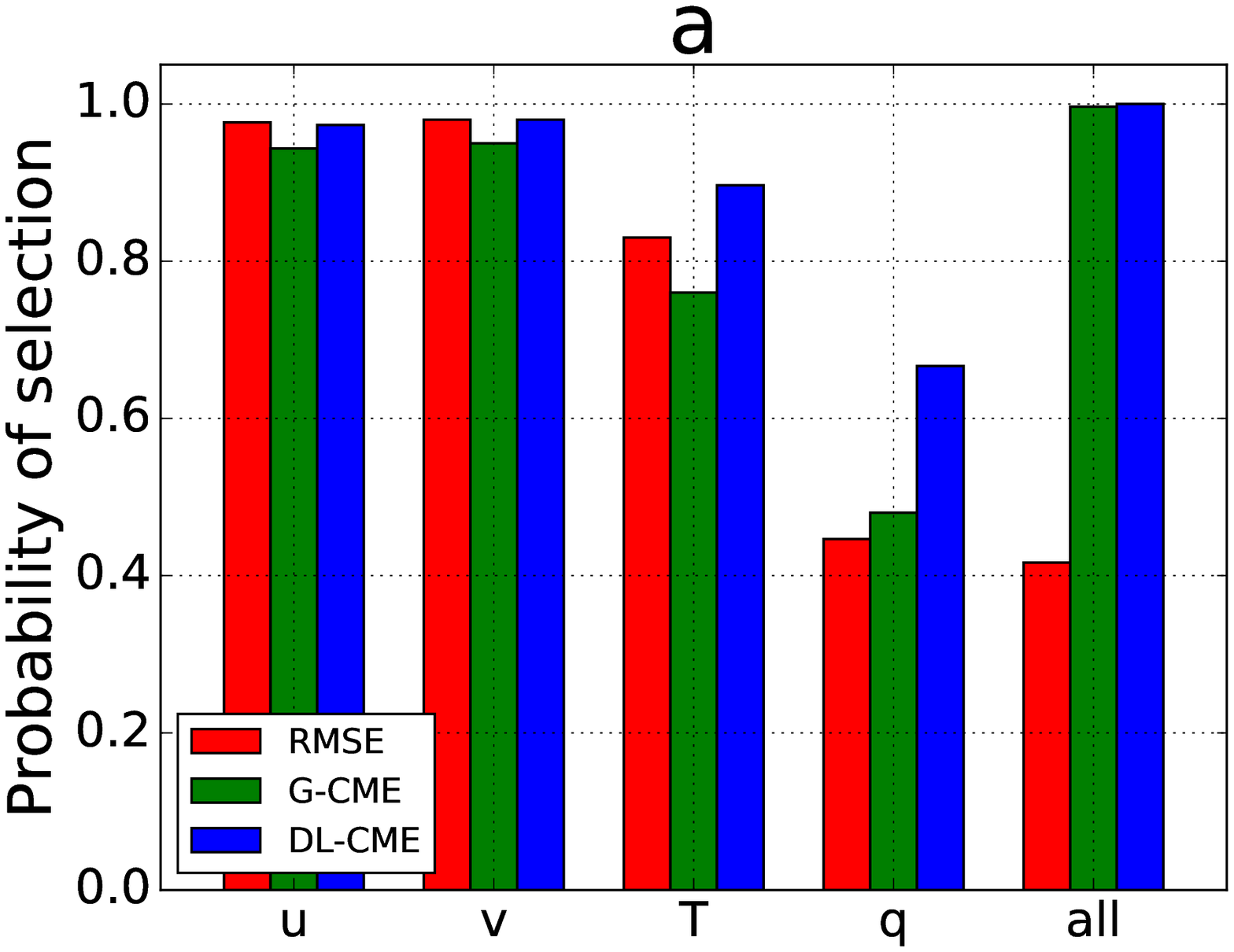} &
						\includegraphics[width=8cm]{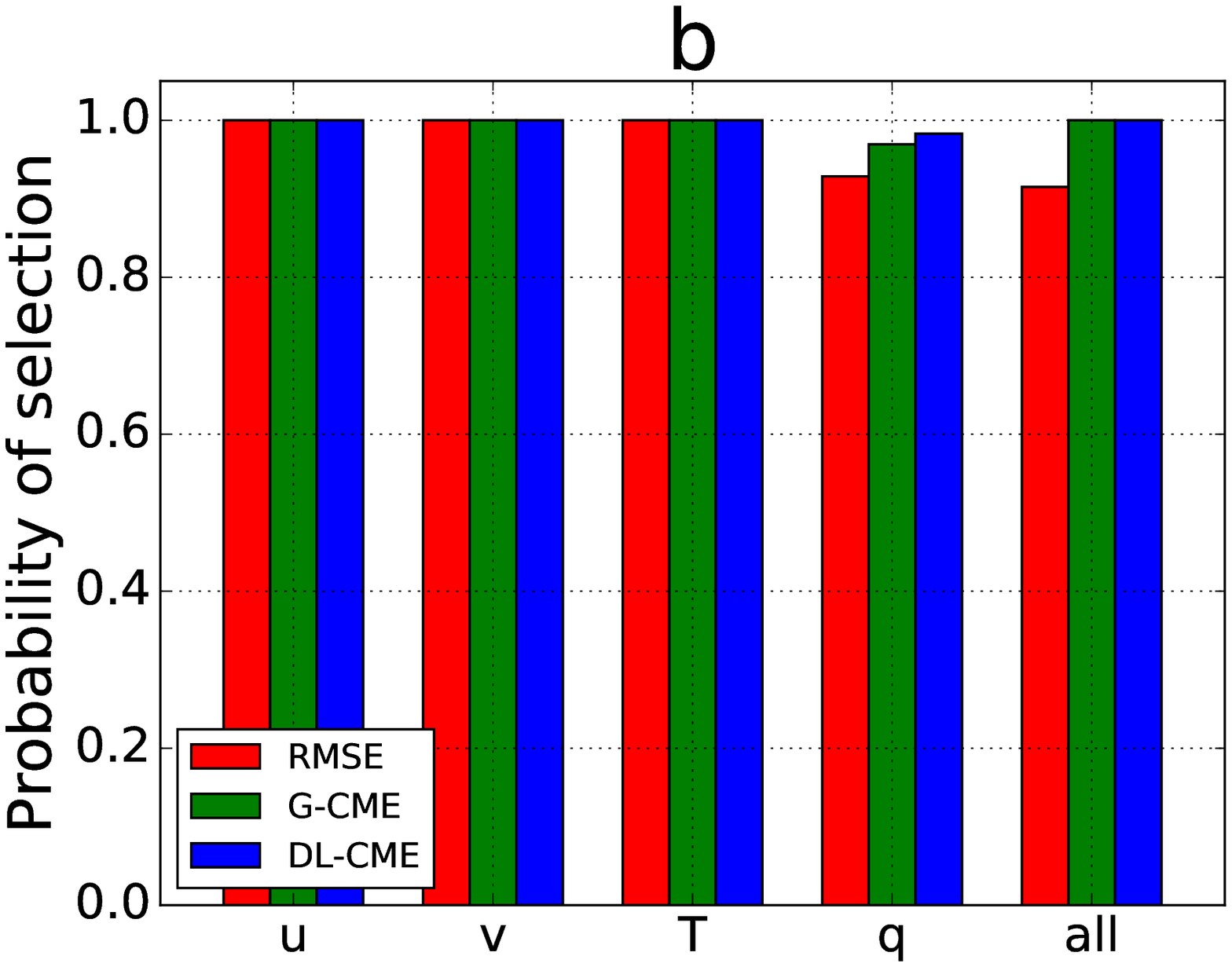}
					\end{tabular}
					
					\caption{
Probability of selection of the correct parameter value over the incorrect one, obtained by using three {\correction model version selection} indicators: RMSE (red bar on the left), the G-CME (green bar in the middle) and the DL-CME (blue bar on the right). The results are given for the variables $u$, $v$, $T$, $q$, and for all variables (`all'); the evidencing window has length (a) $K=1$, and (b) $K=12$. 
						}
					\label{Figure7}
				\end{figure*}
			\end{center} 
			
			 \subsubsection{The model's convection parametrization}
			 \label{conv_par} 
			 SPEEDY's convection scheme is a simplified version of the mass-flux scheme developed by \cite{tiedke93}. 
			 The scheme simulates the balance of the updraft of the saturated air and the large-scale downward motion in between the {planetary boundary layer} (PBL) and the {top-of-convection} (TCN) layer. 
			 While the updraft of saturated air is triggered above the PBL, the detrainment only takes place in the TCN layer. 
			 In addition, the convection scheme simulates an exchange of moisture between the PBL and the layers above it. 
			 
			 SPEEDY's simplified convection parametrization may roughly be summarized by three main steps: an activation of the convection scheme, its closure and an exchange mechanism between the intermediate layers. 
			 In particular, the closure of the scheme requires the convective moisture tendency in the PBL to be equivalent to a relaxation of humidity towards a prescribed threshold with a relaxation time equal to $\tau_\mathrm{cvn}$; in our study, the parameter of interest is precisely $\tau_\mathrm{cnv}$.
			 This parameter has a straightforward physical interpretation since it controls the speed of the vertical mixing associated with moist convection, and its interest lies in its impact on the global--scale circulation. 
			 
			 The correct parameter value,  namely the value set in SPEEDY to create a `true' trajectory --- in the sense defined in sections~\ref{ModEvidAndDA} and \ref{NumExpL95} --- is $\tau_\mathrm{cnv} = 6$ hr. 
			 Meanwhile, the incorrect parametrization here will use the relaxation time $\tau_\mathrm{cnv}=9$ hr.   
			 The objective is to be able to select the correct parameter over the incorrect one, 
			 based on the observation of what is deemed to be the true trajectory in the present, ``dizygotous-twin'' setting; see, for instance, \cite{ghil91} for an explanation of the distinctions between identical-twin, dizygotous-twin and real-observations setup in DA studies.
			 The DA formulation needed to do the parameter selection in this setup is presented in the next subsection.    
 
			 \subsubsection{The DA implementation} 
			 \label{DAimplSPEEDY}
			 As in the L95 experiments of section~\ref{ModEvidAndDA}, a true trajectory $\textbf{x}^\mathrm{t}$ is generated using the correct parameter value of $\tau_\mathrm{cnv}=6$ hr.
			 From this trajectory a set of observations are derived and made available every 6~hr. 
			 This DA configuration was adapted for SPEEDY by \cite{miyoshi05}. 
			 
			 The observed physical quantities are the 5 prognostic variables $u$, $v$, $T$, $q$, throughout the model's 7 layers, and the surface pressure $p_s$. 
			 All variables are observed every four gridpoint, yielding $(96\times 48\times 7) /8=4032$ observations for $u$, $v$, $T$, $q$ and $(96\times 48) /8=576$ observations for the surface pressure.  
			 The observation error variances, used to create the observations in the DA process, are: 
			 \vspace*{-0.7cm}
			 \begin{eqnarray*} 
			 	\sigma_{u} & = & \sigma_{v} = 1 \text{  m.s$^{-1}$, }\\ 
			 	\sigma_{T} & = & 1 \text{ $^{\circ}$C,} \\
			 	\sigma_{q} & = & 1 \text{  g.kg$^{-1}$,} \\
			 	\sigma_{p_s} & = & 1 \text{  hPa.}
			 \end{eqnarray*} 
		     \vspace*{-1cm}		 
		     
		     The experiment lasted 6 months, from 1 January to 30 June 1983. 
		     The results for the spin-up time of $T_\mathrm{ su}=124$ DA cycles (i.e., the 1st month) were discarded, and the diagnostics were calculated for the $T=600$ DA cycles of 1 February through 30 June.
		      
			The DA is performed using the LETKF (cf. section~\ref{LETKFmethodo}) with N=50 ensemble members and with the domain localization introduced by \cite{hunt07}, \cite{miyoshi07}, and \cite{greybush11}: localization domains are selected and the observation error covariance sub-matrix $\textbf{R}|_{s}$ is multiplied, for each $s\in\Gamma$, by the inverse of a smoothing function, as explained in section~\ref{DomLoc}.
			In this study, we use the implementation proposed by \cite{miyoshi07b} and apply as a smoothing function, in the three--dimensional physical space, the Gaussian function 
			 \begin{equation}
			 \phi(r)=\exp \left( \frac{ -r^2 }{2\rho_\mathrm{loc}^2} \right);
			 \end{equation}
			 here $r$ is the distance between the localization domain's center and the observations, while $\rho_\mathrm{loc}$ is the localization radius, and both quantities have a horizontal and a vertical component. 
			 The horizontal component of $\rho_\mathrm{loc}$ is taken equal to 700 km, while its vertical component is defined in terms of log-pressure coordinates and corresponds to approximately 4 km.   
			 
			 Note that, in the following, the DL-CME results are computed for specific variables, i.e., we assume that only these variables have been observed. 
			 This computation is identical to the computation of the CME for an observation subvector, as discussed in section \ref{CMEsubvector}.
			  
			 For an evidencing window that has length $K\ge2$, the required computation of the pdf of the system's state, conditioned on this particular observation subvector leads to a DA run that does not converge. 
			 Hence, we approximate once again this conditional pdf by the original posterior pdf which consists in directly applying the EnKF-formulation to the observation vector reduced to a specific variable using Eq. (\ref{CMEtilde2}).

		\subsection{{\correction Model version selection} problem}      
		We now focus, as explained in section~\ref{conv_par}, on determining which value of the relaxation parameter $\tau_\mathrm{cnv}$ (6 or 9 hrs) is more appropriate to describe the atmospheric evolution simulated during the 5 months of the experiment. 
		Recall that the value $\tau_\mathrm{cnv}=6$ hrs is the one used to simulate the reference truth from where the observations are sampled. 
		The {\correction model version selection} is here a parameter identification problem and, again, is carried out with the RMSE, the G-CME, and the DL-CME as indicators. 
		   
		The indicators are first computed on the evidencing window $K=1$, i.e., on a single DA cycle then averaged over the 5--month interval. 
		The results in terms of selection probability obtained on the various variables are shown in Figure~\ref{Figure7}(a).  
		 
		As mentioned in section~\ref{conv_par}, the value of the relaxation time $\tau_\mathrm{\rm cnv}$ plays an important role in the global-scale circulation: it intensifies, for instance, the mesoscale circulations associated with deep convection which then affects the extent and intensity of the Hadley and Walker cells (see for instance \citealp{donner01}).  
		This physical connection is confirmed by the results of the three {\correction model version selection} indicators when using observations of the variables $u$ or $v$ alone. 
		Indeed, for either of these two variables, the indicators select the correct parameter with high probability.
		
		Note that the RMSE outperforms the G-CME selecting skills on these two variables. 
		However, taking into account localization, i.e., using the DL-CME, improves the selection skills of the CME and provides equally high selection probabilities as the RMSE does. 
		Moreover, when only T-data are used, the DL-CME identifies the correct parameter with the highest probability. 
		The humidity variable, $q$, is the variable least impacted by the variation of $\tau_\mathrm{cnv}$. 
		Yet, the DL-CME is the best to emphasize the improvement of the correct parameter on the humidity representation. 
		Finally, when using all observations, both CMEs fully discriminate between the two {\correction model versions} whereas the RMSE only partially does.   
		
		A more realistic setting is to use a longer evidencing window, taken now as $K = 12$, which corresponds to 3 days.
		The results of this experiment are shown in Figure~\ref{Figure7}(b), and they confirm that increasing $K$ improves the selection skills of all indicators. 
		The humidity is still the most difficult variable to discriminate the two {\correction model versions} from. 
		Yet, the DL-CME remains the best selector, strongly selective for all variables, both separately and together.  
		Moreover, when using all observations, the RMSE fails to systematically recognize the correct {\correction model version} from the incorrect one.  
	 
		From a practical stand-point, the latter result indicates that if, for instance, one's goal is to determine which of two {\correction model versions} is most likely responsible for a three-day weather event that was fully observed, the DL-CME is more likely than the RMSE to provide the correct answer.    
		
		\begin{center}
			\begin{figure*}[h!]
				\centering
				\begin{tabular}{cc} 
					\includegraphics[height=6cm,width=9cm]{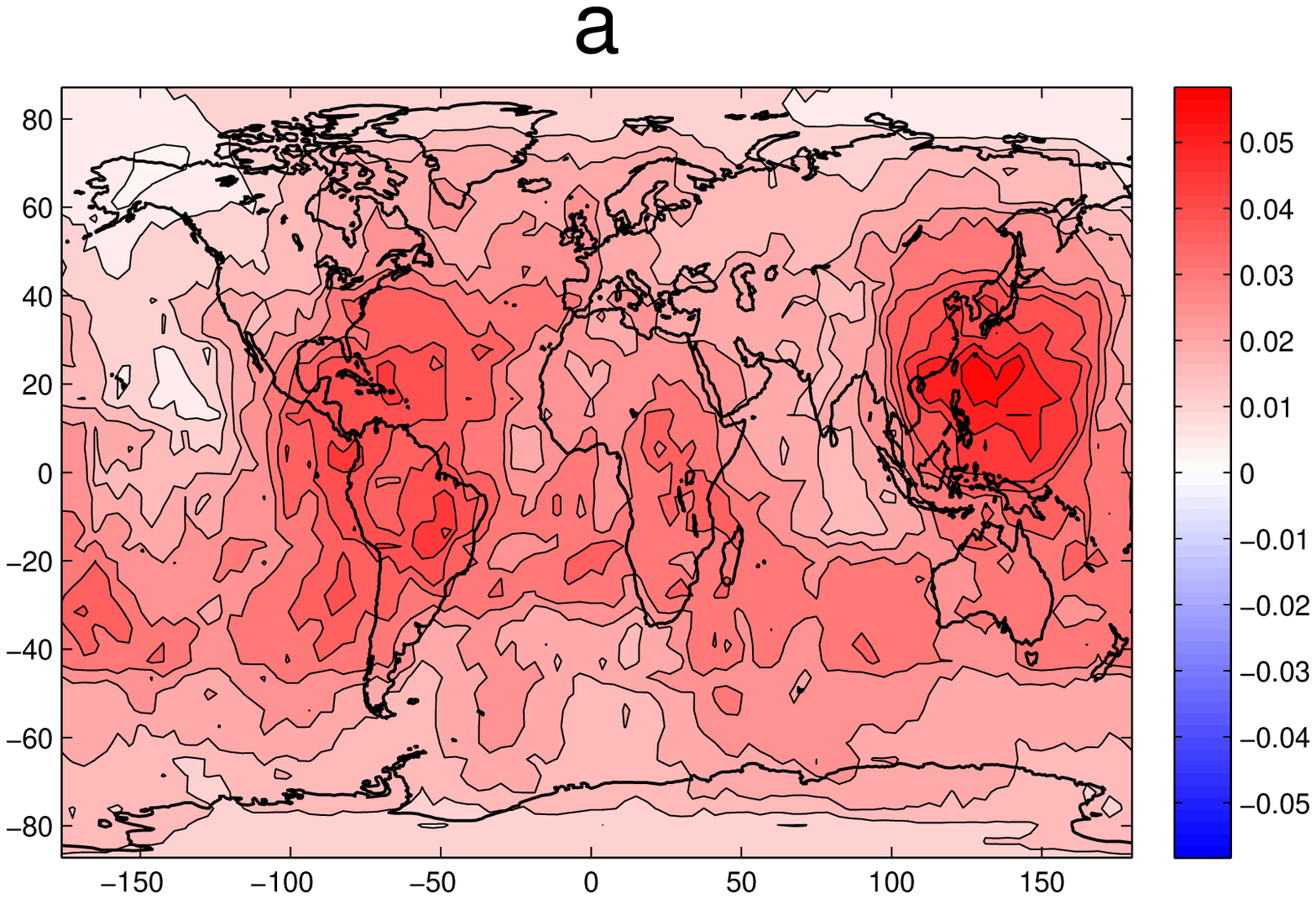} &
					\includegraphics[height=6cm,width=9	cm]{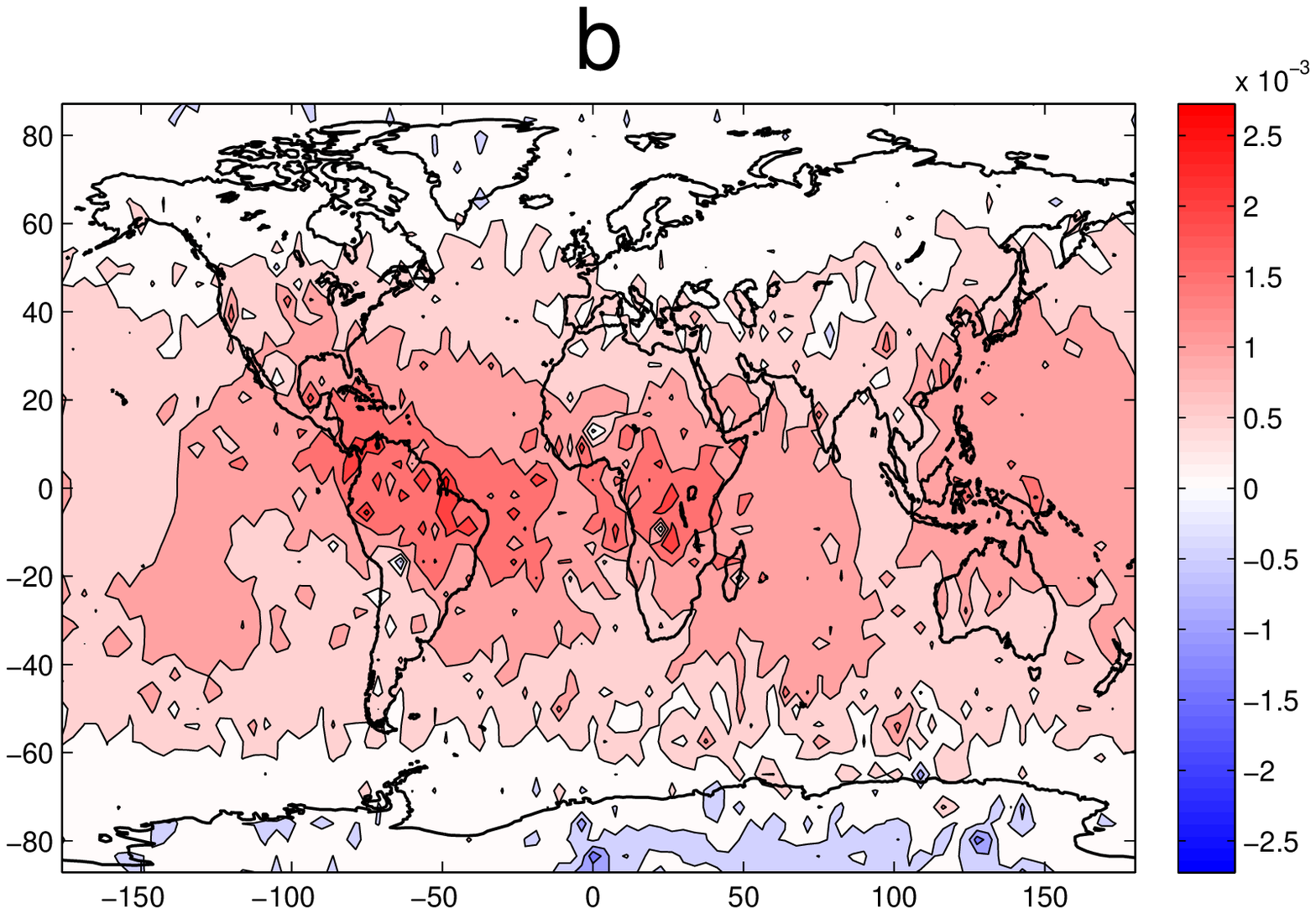} 
				\end{tabular}
				\caption{ 
					Spatial difference between the {\correction model version} with the correct and incorrect parameter values, shown for humidity $q$ and $K=1$, and averaged over the 5-month interval. (a) RMSE results, and (b) DL-CME results. 
					}
				\label{Figure8}
			\end{figure*}
		\end{center}

			\subsection{Mapping model evidence}     
	Local CMEs, defined by Eq. (\ref{CMEtilde2}),  also provide a spatial representation of model evidence.  
	This is exploited in Figure \ref{Figure8} that shows the spatial differences between the correct and the incorrect parametrization using all observations.
		 
	Note that the distinct information conveyed by these two maps --- the local RMSE differences in panel (a) and the local CME differences in panel (b) --- is not necessarily contradictory but rather complementary since the RMSE is the Euclidean distance between the DA forecast and the observations while the local CME can be seen as a distance between the same two quantities but smoothed by the uncertainties that surround them.   
	
	Indeed, the local RMSE reveals a significant difference between the two {\correction model versions} all over the globe while the local CME shows that this difference mainly impacts the low latitude regions. 
	Also, some of the regions that are strongly impacted by the incorrect parameter according to the RMSE, are not so according to the local CME. 
	For instance, the differences emphasized by the RMSE in the region off the coast of China are sensibly smaller according to the local CMEs. 
	This would suggest that, although the DA forecast produced by the incorrect {\correction model version} does not fit the observations as well as the one produced by the correct {\correction model version}, the uncertainties generated in this region mitigate the quality of the correct {\correction model version}'s representation of this region.     


	\section{Concluding remarks}
	\label{RemAndConclu} 
	
	\subsection{Summary and conclusions}
	\cite{carrassi17} have introduced the contextual model evidence (CME) as a powerful indicator for {\correction model version selection} and for attribution of climate related events. 
	They also have provided analytic ways to compute this indicator by using a hierarchy of ensemble-based data assimilation (DA) algorithms and have shown its efficiency.  
	
	When applied to large-dimensional models, ensemble-based DA requires localization techniques in order to mitigate the spurious effect of under-sampling.  
	The present study focused on adapting the CME's EnKF-formulation to the use of a specific localization technique, namely domain localization. 
	
	In section~\ref{CMEsubvector}, we have shown how to compute local CMEs as the CMEs of an observation subvector. 
	This idea led to an EnKF-formulation of the local CME, given by Eq. (\ref{CMEtilde2}), subject to very few simplifying assumptions. 
	Combining local CMEs for each grid point, we introduced in Eq. (\ref{DomainLikeliEq}) a heuristic {\correction model version selection} indicator called the domain-localized CME (DL-CME). 
	This DL-CME formulation allows one to estimate model evidence in a consistent and routine way by using localized DA with large-dimensional problems. 
		
		We implemented the DL-CME formulation --- along with the global CME's EnKF-formulation called global CME (G-CME) and the root mean square error (RMSE), defined by Eq. (\ref{RMSE}) --- on two atmospheric models: the 40-variable mid-latitude atmospheric dynamics model (L95: \citealp{lorenz98}) and the simplified global atmospheric SPEEDY model \citep{molteni03}, in sections~\ref{NumExpL95} and \ref{NumExpSPEEDY}, respectively.
		The main objective of the numerical experiments in these two sections was to assess the performance of the three approaches so defined --- the RMSE, the G-CME and the DL-CME --- in increasingly harder {\correction model version selection} problems.  
		  
	In section \ref{NumExpL95}, the numerical experiments based on the L95 model were used to compare the three {\correction model version selection} indicators for varying model forcings $F$, as well as with respect to the localization radius $\rho_\mathrm{loc}$ and to the length $K$ of the evidencing window. 
	In general, both CME indicators, G-CME and DL-CME, are more accurate than the RMSE; see Figures~\ref{Figure1}--\ref{Figure6}.
	Moreover, the DL-CME outperforms systematically the G-CME regardless of the length of the localization radius, cf. Figure~\ref{Figure5}. 
		
	In the 40-ensemble member localization-free experiment, the G-CME increasingly outperforms the RMSE when the difference between the model forcings is increased; see Figures~\ref{Figure1}-\ref{Figure2}. 
	When the number of ensemble members is reduced to 10, hence making localization necessary, the G-CME is still more accurate than the RMSE. 
	With the same ensemble size, the DL-CME is even more accurate than the G-CME which indicates the impact of consistently using localization in both the DA state estimation procedure and in the CME computation. 
	Remarkably, DL-CME, with a 10-member ensemble and localization, produces better selection probabilities than G-CME with a 40-member ensemble and no localization, at least when the discrepancy between the correct and incorrect forcing is large enough, cf. Figures~\ref{Figure3}-\ref{Figure4}. 
	 
	The sensitivity of the {\correction model version selection} skill to the localization radius is approximately the same for all indicators, meaning that the improvement of the DL-CME over the two other indicators is almost independent of the localization radius, cf. Figure~\ref{Figure5} . 
	This independence is highly desirable, given the computational cost of tuning the radius.   
	In particular, the DL- CME is clearly the most accurate indicator for small evidencing windows. 
	For larger windows, the performances of the DL-CME and the G-CME become almost identical and do remain higher than the RMSE's performance, see Figure~\ref{Figure6}.

	In section \ref{NumExpSPEEDY}, {\correction model version selection} was carried out for a more realistic problem. 
	We formulated two parametrizations of the convection in the SPEEDY model by varying the convective relaxation time $\tau_\mathrm{cnv}$.  
	The selecting skills of the RMSE, the G-CME and the DL-CME were assessed on all physical model variables, both separately and together.   
	
	Over a single DA-cycle, the DL-CME and the G-CME are fully selective on all variables combined. 
	The DL-CME, though, is the best selection indicator on all four model variables --- $u$, $v$, $T$ and $q$ --- separately, whether for a single DA-cycle or a large evidencing window, $K = 12$, as shown in Figure~\ref{Figure7}. 
 
	Local CMEs provide, in addition, a spatial view of model evidence, cf. Figure~\ref{Figure8}. 
	This spatial view can provide precious information on the impacts of the correct and incorrect parametrizations on the physical evolution of the large-scale atmospheric flow. 
	 
	In summary, while the DL-CME displays the best performance in terms of {\correction model version selection}, the spatial plots of local CME can help identifying the spatial regions and the physical variables where the incorrect {\correction model version} misfits the observations.  
	Beyond {\correction model version selection} problems, this feature can be used to bring precious information in order to help understanding the causal chain leading to meteorological or climatological events as well as designing target observations/areas for that purpose. 	
	
	\subsection{Future directions} 
	 
	A first line of future investigations should focus on strengthening the theoretical foundation of the CME. 
	Indeed, several theoretical issues are yet to be addressed. 
	For instance, simulating model error and taking it into account in both the DA process and in the CME computation could improve significantly the CME's {\correction model version selection} skill.
	The DA community is still conducting active research on the best ways to incorporate model error in DA methodology (e.g., \citealp{dee95,raaenes15,carrassi16,harlim17}). 
	This research has the potential to improve significantly the accuracy of the DA product improving therewith the CME estimation. 
	Moreover, taking into account model error in the CME computation itself could also have an impact in the CME's {\correction model version selection} abilities. 
	
	{\correction
		A second potential limitation of the CME method is the accuracy of the CME estimation when computed with a sparsely observed system.  
		This limitation obviously depends on the accuracy of the assimilation itself. 
		However, as long as the assimilation is possible {\correction -- i.e. as long as the spatial and temporal density of the observations is sufficient for the system to be observable -- and the two compared model versions have the same} observation system, one can assume that they will be similarly impacted. 
		A study of the observation system's impact on the {\correction CME's model version selection skill} and, more generally, on the CME estimation should be conducted {\correction in the future.}.
		}
	
	Another theoretical issue requiring investigation is the derivation of an exact global indicator based on local CMEs that could outperform the global heuristic indicator proposed in this article, the DL-CME. 
	These theoretical challenges are part of the authors' current research program. 
	 
	A second line of investigations should focus on broadening the applications of the CME approach.
	{\correction
		The realism and the complexity of the models considered {\correction needs to be} increased. 
		Indeed, the CME approach allows one to compare model {\correction versions} that differ from one another {\correction by the values of} a set of parameters. 
		
		In the numerical experiments presented here, the {\correction model versions differed only by a parameter's} scalar value, yet one {\correction may wish to} consider two complex model {\correction versions} that differ by {\correction by the values of} a large set of parameters or by an entire two-dimensional forcing field, {\correction like the sea surface temperatures for an atmospheric model.
		For instance, in currently} ongoing work, the authors are using the CMEs to identify the best turbulence closures among a number of full parametrization schemes. 
		}
		
	 {\correction
		  At the same time, {\correction application of the CME approach should be widened from model version selection} to other purposes, such as {\correction causal attribution of observed changes.}
		}
	In particular, it seems worth investigating the potential of the spatial representation of local CMEs for diagnosing causality in a geoscientific system, as proposed by \cite{hannart16b}.  

		
	\section*{Acknowledgments}
	We acknowledge discussions with the other members of the DADA team (Philippe Naveau, Aur\'elien Ribes and Manuel Pulido). 
	We also wish to acknowledge personal communications with Pr. P.J. Van Leeuwen which led to clarifications of the manuscript.
	 
	This research was supported by grant DADA from the Agence Nationale de la Recherche (ANR, France: All authors), and by Grant N00014-16-1-2073 of the Multidisciplinary University Research Initiative (MURI) of the US Office of Naval Research (MG). 
	A. Carrassi has been funded by the project REDDA of the Norwegian Research Council (contract number 250711).
	J. Ruiz has been funded by the project PICT2014-1000 of the Argentinian National Agency for Scientific Research Promotion. 
	CEREA is a member of the Institute Pierre Simon Laplace (IPSL). 

	  
	\begin{appendices} 
	\section[Appendix A]{\\Alternative implementation of the global CME (G-CME)}
	\label{Appendix1}   
	The G-CME implementation was first presented in \cite{carrassi17} for the CME formulation based on the En-4D-Var DA method.  
	This implementation makes the assumption that the observation error covariance matrix $\textbf{R}$ is diagonal.  
	This assumption is strong but often verified in geosciences applications. 
	
	Implementing Eq. (\ref{EnKFLikelihoodFormulation}) requires to derive: (i) the determinant term $| \boldsymbol{\Sigma}_k|$ ; (ii) the weighted sum of squares $(\textbf{y}_k - \textbf{H}_k\textbf{x}^\mathrm{f}_k )^\mathrm{T}\boldsymbol{\Sigma}^{-1}_k(\textbf{y}_k - \textbf{H}_k\textbf{x}^\mathrm{f}_k )$. 
	We deal with each two terms separately.
	
	First, the computation of the determinant term can be greatly simplified by application of Sylvester's determinant rule (i.e., $| \textbf{I} + \textbf{AB}| = | \textbf{I} + \textbf{BA}|$ for any two rectangular matrices $\textbf{A}$ and $\textbf{B}$ of conformable sizes). 
	We have 
	\begin{align}
	\begin{split}
		| \boldsymbol{\Sigma}_k|
		&= | \textbf{H}_k\textbf{P}^\mathrm{f}_k\textbf{H}_k^\mathrm{T}+\textbf{R}| \\
		&=| \textbf{R}| . | \textbf{I}_d+ \textbf{R}^{-\frac{1}{2}}\textbf{H}_k\textbf{P}^\mathrm{f}_k\textbf{H}_k^\mathrm{T}\textbf{R}^{-\frac{1}{2}}|\\
		&=| \textbf{R}| . | \textbf{I}_N+\boldsymbol{\Delta}_k| 
		\end{split}
		\end{align}
		where $\boldsymbol{\Delta}_k=(\textbf{X}^\mathrm{f}_k)^\mathrm{T}\textbf{H}_k^\mathrm{T}\textbf{R}\textbf{H}_k\textbf{X}^\mathrm{f}_k$.
		with $\textbf{X}^\mathrm{f}_k$, the normalized forecast anomalies such that $\textbf{X}^\mathrm{f}_k(\textbf{X}^\mathrm{f}_k)^\mathrm{T}=\textbf{P}^\mathrm{f}_k$. 
		Note that $\textbf{R}$ is diagonal in general thus its determinant is easy to compute.

		Under Eq. (2), one sees that $| \boldsymbol{\Sigma}_k|$ boils down (after the one off computation of $| \textbf{R}|$) to the computation of $| \textbf{I}_N + \boldsymbol{\Delta}_k |$ which is a much smaller determinant of dimension $N$ (size of the ensemble) instead of dimension $d$ (size of the observation vector).
		
		Second, the computation of the weighted sum of squares can also be greatly simplified by application of the Sherman-Morrison-Woodbury formula. 
		We have 
		\begin{align}
		\begin{split}
			\boldsymbol{\Sigma}^{-1}_k
			&=  (\textbf{H}_k\textbf{P}^\mathrm{f}_k\textbf{H}_k^\mathrm{T}+\textbf{R})^{-1} \\ 
			&=  (\textbf{H}_k\textbf{X}^\mathrm{f}_k(\textbf{X}^\mathrm{f}_k)^\mathrm{T}\textbf{H}_k^\mathrm{T}+\textbf{R})^{-1} \\
			&= \textbf{R}^{-1} -\textbf{R}^{-1} \textbf{H}_k\textbf{X}^\mathrm{f}_k(\textbf{I}_N+\boldsymbol{\Delta}_k)^{-1}(\textbf{X}^\mathrm{f}_k)^\mathrm{T}\textbf{H}_k^\mathrm{T}\textbf{R}^{-1}
			\end{split}
			\label{Eq3}
			\end{align}
			
			It follows 
			\begin{align}
			\begin{split}
				& (\textbf{y}_k - \textbf{H}_k\textbf{x}^\mathrm{f}_k )^\mathrm{T}\boldsymbol{\Sigma}^{-1}_k(\textbf{y}_k - \textbf{H}_k\textbf{x}^\mathrm{f}_k ) \\
				&= (\textbf{y}_k - \textbf{H}_k\textbf{x}^\mathrm{f}_k )^\mathrm{T} \\ &(\textbf{R}^{-1} -\textbf{R}^{-1} \textbf{H}_k\textbf{X}^\mathrm{f}_k(\textbf{I}_N+\boldsymbol{\Delta}_k)^{-1}(\textbf{X}^\mathrm{f}_k)^\mathrm{T}\textbf{H}_k^\mathrm{T}\textbf{R}^{-1}) \\ &(\textbf{y}_k - \textbf{H}_k\textbf{x}^\mathrm{f}_k ) \\
				&= (\textbf{y}_k - \textbf{H}_k\textbf{x}^\mathrm{f}_k )^\mathrm{T}\textbf{R}^{-1}(\textbf{y}_k - \textbf{H}_k\textbf{x}^\mathrm{f}_k ) -\textbf{v}_k(\textbf{I}_N+\boldsymbol{\Delta}_k)^{-1}\textbf{v}_k^\mathrm{T}
				\end{split}
				\label{Eq4}
				\end{align}
				with $\textbf{v}_k=\textbf{R}^{-1} \textbf{H}_k\textbf{X}^\mathrm{f}_k$. 
				
				Under Eq. (\ref{Eq3}) and (\ref{Eq4}), one sees that the second term boils down (after the one off inversion of $\textbf{R}$) to the inversion of $\textbf{I}_N + \Delta_k$, which again is a much smaller matrix of dimension $N$ instead of dimension $d$, and to the computation of the intermediate vector $\textbf{v}_k$ of dimension $N$ .
				
				\section[Appendix B]{\\Derivation of the CME formulation using {\correction background} covariance localization}
				\label{Appendix2}    
					 {\correction Background} covariance localization (also called Schur localization) consists in restraining the spatial impact of the empirical forecast error covariances.  
					Indeed, the ensemble size is smaller than the model's state dimension and thus the ensemble-based covariance is subsampled by construction. 
					This subsampling may lead to spurious correlations. 
					This issue can be partly overcome by removing the unrealistic correlations. 
					 {\correction Background} covariance localization performs this removal by applying to the forecast error covariance matrix $\textbf{P}_f$ a Schur product (i.e., a pointwise multiplication) with a smoothing correlation matrix $\mathbf{C}$ 
					\begin{equation}
					[\mathbf{C}\circ \textbf{P}_f]_{i,j}=[\textbf{P}_f]_{i,j}[\mathbf{C}]_{i,j}
					\end{equation} 
					where the correlation matrix coefficients are defined by $[\mathbf{C}]_{i,j}=\phi(|i-j|)$, for $\phi$ a smoothing function. 
					 
					The smoothing function we use is a short-ranged predefined correlation function $\phi(r)=G(r/\rho_\mathrm{loc})$, based on the Gaspari-Cohn function \citep{gaspari99}, for the localization radius $\rho_\mathrm{loc}$, 
					\begin{align}
					&\!\!G(z)=\nonumber\\
					& \!\!\left\{ \!\!\! 
					\begin{array}{l}
					\text{if \ } 0\! \leq\! r\! <\! 1\! :\, 1\!-\!\frac{5}{3}z^2\! +\! \frac{5}{8}z^3\! +\!\frac{1}{2}z^4\! -\!\frac{1}{4}z^5, \\
					\text{if \ } 1\!\leq\! r\! <\! 2\! :\, 4\! -\! 5z\! +\! \frac{5}{3}z^2\! +\!\frac{5}{8}z^3\! -\!\frac{1}{2}z^4\! +\!\frac{1}{12}z^5\! -\!\frac{2}{3z}, \\
					\text{if \ }\! r\! \geq 2:\, 0.
					\end{array}
					\right.
					\label{GaspariAndCohn}
					\end{align} 
					
				Based on the {\correction background} covariance localization, a new CME formulation can be derived. 
				We call this formulation: the {Global Localized} CME (GL-CME).
				The GL-CME respects the original formulation Eq. (\ref{EnKFLikelihoodFormulation}), i.e., $f(\textbf{y})$ is computed for the entire observation vector, while taking into account the regularized forecast error covariance matrix in the term $\boldsymbol{\Sigma}_k$ 
				\begin{equation}
				\boldsymbol{\Sigma}_k=\textbf{H}_k[\mathbf{C}\circ \textbf{P}^\mathrm{f}_k] \textbf{H}_k^\mathrm{T} + \textbf{R}.
				\end{equation}
				
				In this case, the CME computation remains quite straightforward. 
				However, concerning the computational cost, the regularization does not change the size of $\textbf{P}^\mathrm{f}_k$. 
				The computational cost is as heavy as the original {EnKF-formulation} which adds to the cost of the application of a $M\times M$ Schur product, making this formulation not directly tractable in high dimensions. 
				To do so, additional sophisticated techniques are required.

\end{appendices} 

\end{document}